\documentclass[12pt,preprint]{aastex}

\shorttitle{Disks Around Young Gas Giant Planets}
\shortauthors{Turner, Lee \& Sano}

\begin{document}
\title{Magnetic Coupling in the Disks Around Young Gas Giant Planets}

\author{N.~J.\ Turner\altaffilmark{1}, Man Hoi Lee\altaffilmark{2} and
  T.\ Sano\altaffilmark{3} }

  \altaffiltext{1}{Jet Propulsion Laboratory, California Institute of
    Technology, Pasadena, California 91109, USA;
    neal.turner@jpl.nasa.gov}
  \altaffiltext{2}{Department of Earth Sciences and Department of
    Physics, The University of Hong Kong, Pokfulam Road, Hong Kong;
    mhlee@hku.hk}
  \altaffiltext{3}{Institute of Laser Engineering, Osaka University,
    Suita, Osaka 565-0871, Japan; sano@ile.osaka-u.ac.jp}

\begin{abstract}
  We examine the conditions under which the disks of gas and dust
  orbiting young gas giant planets are sufficiently conducting to
  experience turbulence driven by the magneto-rotational instability.
  By modeling the ionization and conductivity in the disk around
  proto-Jupiter, we find that turbulence is possible if the X-rays
  emitted near the Sun reach the planet's vicinity and either (1) the
  gas surface densities are in the range of the minimum-mass models
  constructed by augmenting Jupiter's satellites to Solar composition,
  while dust is depleted from the disk atmosphere, or (2) the surface
  densities are much less, and in the range of gas-starved models fed
  with material from the Solar nebula, but not so low that ambipolar
  diffusion decouples the neutral gas from the plasma.  The results
  lend support to both minimum-mass and gas-starved models of the
  protojovian disk: (1) The dusty minimum-mass models have internal
  conductivities low enough to prevent angular momentum transfer by
  magnetic forces, as required for the material to remain in place
  while the satellites form.  (2) The gas-starved models have
  magnetically-active surface layers and a decoupled interior ``dead
  zone''.  Similar active layers in the Solar nebula yield accretion
  stresses in the range assumed in constructing the circumjovian
  gas-starved models.  Our results also point to aspects of both
  classes of models that can be further developed.  Non-turbulent
  minimum-mass models will lose dust from their atmospheres by
  settling, enabling gas to accrete through a thin surface layer.  For
  the gas-starved models it is crucial to learn whether enough stellar
  X-ray and ultraviolet photons reach the circumjovian disk.
  Additionally the stress-to-pressure ratio ought to increase with
  distance from the planet, likely leading to episodic accretion
  outbursts.
\end{abstract}


%%%%%%%%%%%%%%%%%%%%%%%%%%%%%%%%%%%%%%%%%%%%%%%%%%%%%%%%%%%%%%%%%%%%%%%%%%%%%%%
\section{INTRODUCTION\label{sec:intro}}

Jupiter's regular satellites have nearly coplanar orbits with small
eccentricities, and probably originated in an orbiting circumplanetary
disk of dust and gas --- a Solar nebula in miniature
\citep{1982Icar...52...14L}.  As Jupiter approached its present mass,
its tides opened a gap in the solar nebula \citep{1986ApJ...307..395L,
  1993prpl.conf..749L}.  Incoming gas then had too much angular
momentum to fall directly onto the planet, and instead went into
orbit, forming a circumjovian disk \citep{1999ApJ...526.1001L}.  The
disk governed the flow of material to the planet and provided the
environment in which the satellites formed.  A key question is
therefore how quickly the orbital angular momentum was redistributed
within the disk, allowing some material to accrete on the planet and
some to spiral outward where it may have been removed by Solar gravity
or by photoevaporation.  Also, was the flow laminar or turbulent?  Did
the released gravitational potential energy become heat in the
interior, or was it dissipated in the disk atmosphere?  And what did
the resulting internal temperatures, densities and flow fields mean
for the processing of the moon-forming ices and silicates?

As with the much larger disks orbiting young stars
\citep{1974MNRAS.168..603L, 1981ARA&A..19..137P, 1995ARA&A..33..199B,
  2011ARA&A..49..195A}, circumplanetary disks' evolution is controlled
by the transport of orbital angular momentum.  In other astrophysical
disks, magnetic forces carry angular momentum outward in the
turbulence resulting from magneto-rotational instability or MRI
\citep{1991ApJ...376..214B, 1998RvMP...70....1B}.  The instability can
work only if the disk material is ionized enough to couple to the
magnetic fields.  In the disks around young giant planets, as in
protostellar disks, the low temperatures mean thermal ionization is
ineffective except very near the central body.  Ionization by
radioactive isotopes' decay, lightning, bolide impacts and
planetesimal ablation is also weak \citep{1996Icar..123..404T}.
Adequate ionization might be produced by interstellar cosmic rays if
not for rapid recombination on the surfaces of dust grains
\citep{2011ApJ...743...53F}.  Our purpose here is to find whether the
disk around Jupiter is ionized enough for MRI turbulence if an
additional ionization process is considered: the X-rays from the young
Sun \citep{1999ApJ...518..848I}.  Below we compute the magnetic
coupling, which depends on the ionization state, which in turn depends
on the distribution of densities and temperatures.

A variety of models has been proposed for the circumjovian disk.  We
consider typical examples from two broad classes.  In the minimum-mass
models \citep{1982Icar...52...14L, 2003Icar..163..198M}, all the
ingredients for the satellites are present from an early stage.  The
gases are eventually dispersed while all the solids are incorporated
into the satellites.  The disk surface density, obtained by augmenting
the rock and ice of the Galilean satellites with gases to Solar or
near-Solar composition, is about $10^7$~g~cm$^{-2}$ at the surface of
the planet with a power-law radial falloff.  The large mass column
means few cosmic rays or X-rays penetrate the interior.  Recombination
is rapid, and the minimum-mass disk couples poorly to magnetic fields
\citep{1996Icar..123..404T}.

The second class of models is gas-starved \citep{2002AJ....124.3404C,
  2006Natur.441..834C}.  Gas and dust trickle into the disk from the
surrounding Solar nebula.  While some of the solids accumulate into
larger solid bodies, much material is lost to the planet through the
effective viscosity of the gas and the gravitational torques exerted
by the gas on the proto-satellites.  In this picture, today's moons
are the last generation to form before the gas dispersed.  The disk
surface densities in this model are less than $1\,000$~g~cm$^{-2}$,
low enough that some cosmic rays can reach the midplane
\citep{2011ApJ...743...53F}.

While disks of gas and dust have been used to explain the moons of
both Jupiter and Saturn \citep{2010ApJ...714.1052S}, other classes of
model may be required given that Jupiter has four large satellites
with a gradient in density, while Saturn has just one large satellite.
Saturn's smaller inner moons may have grown from ring particles
transported out across the Roche limit, with the more distant
experiencing more mergers \citep{2012Sci...338.1196C}.  However it is
unclear whether Titan formed the same way.  Another mechanism,
gas-poor planetesimal capture \citep{1986sate.conf...89S,
  2006Icar..181..486E}, involves collisions among a swarm of
planetesimals.  This picture has not so far yielded a quantitative
accounting for the Galilean moons' large masses and their decrease in
density with distance from the planet.  We therefore focus on a
circumplanetary gas and dust disk as the most promising model for the
origins of Jupiter's large moons.

The paper is laid out as follows.  The minimum-mass circumplanetary
disk models are described in section~\ref{sec:mmcjd}, the gas-starved
models in section~\ref{sec:gssn}.  The chemical reaction network used
to compute the magnetic diffusivities is laid out in
section~\ref{sec:ionization} and the MRI turbulence criteria in
section~\ref{sec:mri}.  The resulting distributions of magnetic
activity in the disks are shown in section~\ref{sec:deadzones}.
Implications for the evolution of the dust and the growth of
satellites are discussed in section~\ref{sec:solids}, and our
conclusions are presented in section~\ref{sec:conclusions}.

\section{MINIMUM MASS CIRCUMJOVIAN DISK MODELS
  \label{sec:mmcjd}}

The minimum-mass models of the circumjovian disk are built in a
similar way to minimum-mass Solar nebula models.  The satellite
system's mass of $2.1\times10^{-4} M_J$ is combined with enough
hydrogen and helium to reach Solar composition.  The resulting disk
has a few percent of Jupiter's mass $M_J$, and extends from inside the
present orbit of Io at $5.9 R_J$ to at least the orbit of Callisto at
$26 R_J$ (where $R_J$ is the radius of Jupiter).

Temperatures in the disk's outer reaches must remain below the water
sublimation threshold to account for the ice-rich makeup of Ganymede
and Callisto.  The release of gravitational energy as disk material
spirals toward the planet may raise temperatures too high unless the
accretion-stress-to-gas-pressure ratio $\alpha<10^{-5}$
\citep{2003Icar..163..198M}.  Stresses near or above this danger level
potentially arise from the damping of the wakes raised in the disk gas
by satellitesimals \citep{2001ApJ...552..793G} and from the stellar
tides periodically forcing the disk \citep{2012A&A...548A.116R}.
However in this paper we focus on whether magnetic forces can yield
still larger stresses.

Another constraint comes from observing that Callisto appears to be
only partly differentiated \citep[moment of inertia $I/MR^2 \approx
  0.355$;][]{2001Icar..153..157A} though we note that it would be
desirable to have the partly-differentiated interpretation confirmed
\citep{1997Icar..130..540M, 2013Icar..226.1185G}.  Keeping ice and
rock mixed is feasible only if the ice never melted during the moon's
assembly.  The gravitational potential energy of the component parts
must then have been released as heat over a period of 0.6~Myr or
longer \citep{2008Icar..198..163B}.  To slow Callisto's growth, it may
be helpful to drop the circumjovian disk's surface density sharply
between Ganymede and Callisto \citep{2003Icar..163..198M}.

Finally, to avoid its ice melting in the heat released by short-lived
radionuclide decay, Callisto also must have finished accreting at
least 4~Myr after the formation of the refractory
calcium-aluminum-rich inclusions \citep{2008Icar..198..163B}.  The raw
materials must persist in orbit around Jupiter until at least this
date.

Each circumjovian disk model is specified by the radial profiles of
gas surface density $\Sigma(r)$, solids-to-gas mass ratio $\phi(r)$
and midplane temperature $T_c(r)$.  From these we obtain the scale
height and density using
\begin{equation}\label{eq:scaleheight}
  H(r) = c_s/\Omega = \left({\cal R} T_c r^3 \over \mu G M_J\right)^{1/2}
\end{equation}
and
\begin{equation}
  \rho(r,z) = {\Sigma(r) \over \sqrt{2\pi} H} \exp[-z^2/(2 H^2)],
\end{equation}
where $c_s$ is the isothermal sound speed, $\Omega$ the orbital
frequency, ${\cal R}$ the gas constant and $\mu=2.3$ the mean
molecular weight, and the density varies with the cylindrical
coordinates $(r, z)$.

For each model we consider versions in which the solids (1) take the
form of sub-micron dust grains, and (2) are locked up in bodies of
1~cm or larger.  Particles this big are few enough that their combined
cross-section for recombination is too low to affect the abundances of
free charges.  We set the sub-micron grains' dust-to-gas mass ratio
$\epsilon(r)$ equal to $\phi(r)$ in the first, dusty case and zero in
the second, dust-free case.

\subsection{Takata \& Stevenson (1996) Model --- MM96}

We include a minimum-mass model very similar to that used by
\cite{1996Icar..123..404T} to facilitate comparison with their
ionization results.  This model, which we call MM96, has the simple
surface density profile
\begin{equation}
\Sigma(r) = \Sigma_0  (R_J/r)
\end{equation}
with $\Sigma_0 = 10^7$\,g\,cm$^{-2}$, and the temperature profile
\begin{eqnarray}
T_c(r) &=& 3600 (R_J/r)\, {\rm K} \qquad {\rm for}\ r/R_J \le 30 \\
       &=& 120\, {\rm K} \qquad {\rm for}\ r/R_J \ge 30.
\end{eqnarray}
The model differs from \cite{1996Icar..123..404T} in that we compute
the density scale height from the temperature via
eq.~\ref{eq:scaleheight}, yielding $H=0.086 r$ within $30R_J$ and
$H\sim r^{3/2}$ beyond, while they simply took $\sqrt{2} H\approx 0.1
r$.  We have checked that the two density distributions yield similar
magnetic diffusivities under X-ray ionization and dust surface
recombination.  Our MM96 model includes a 1\% mass fraction of solid
material.

\begin{figure}[htb!]
\epsscale{0.54}
\plotone{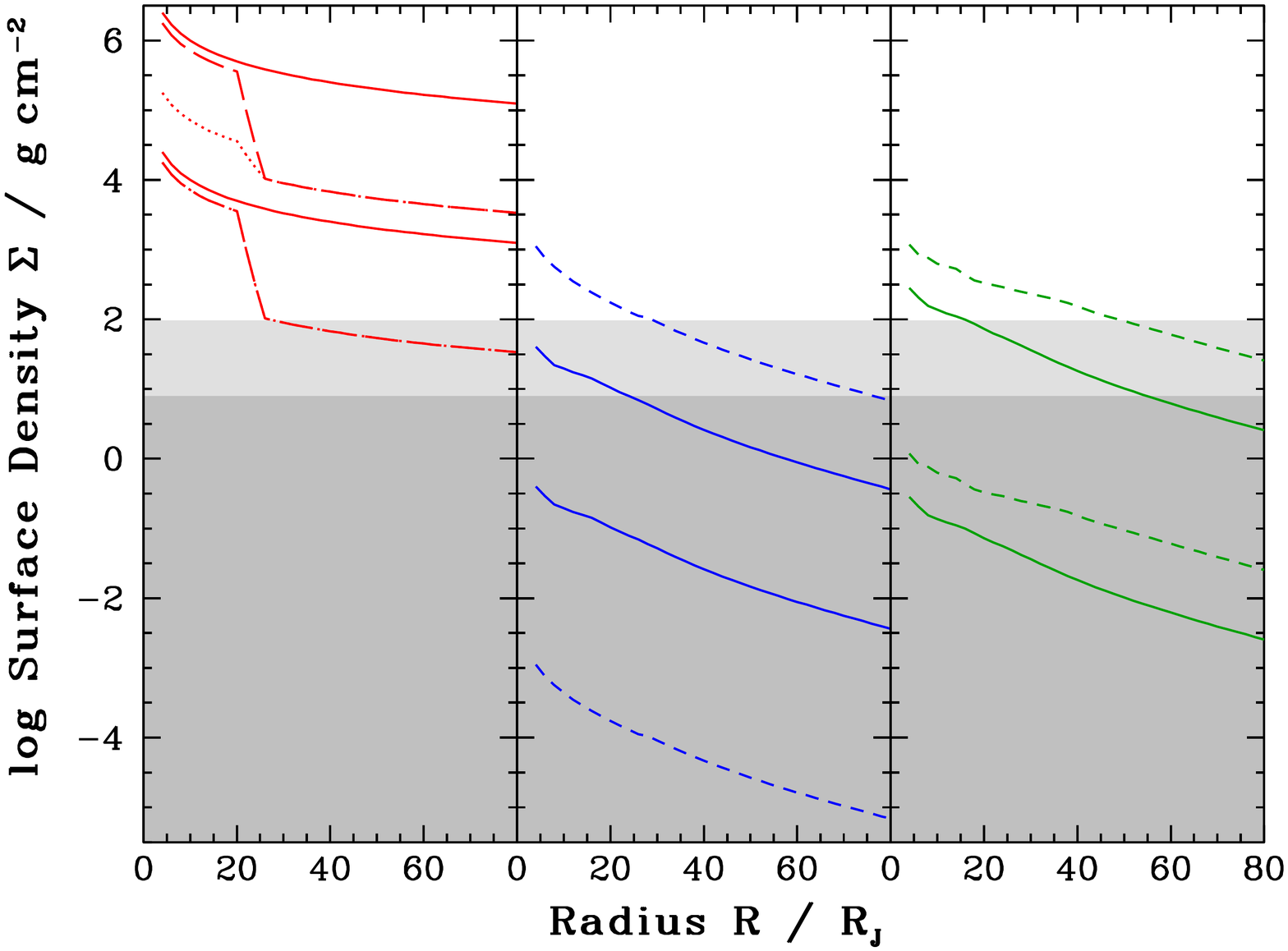}\\
\plotone{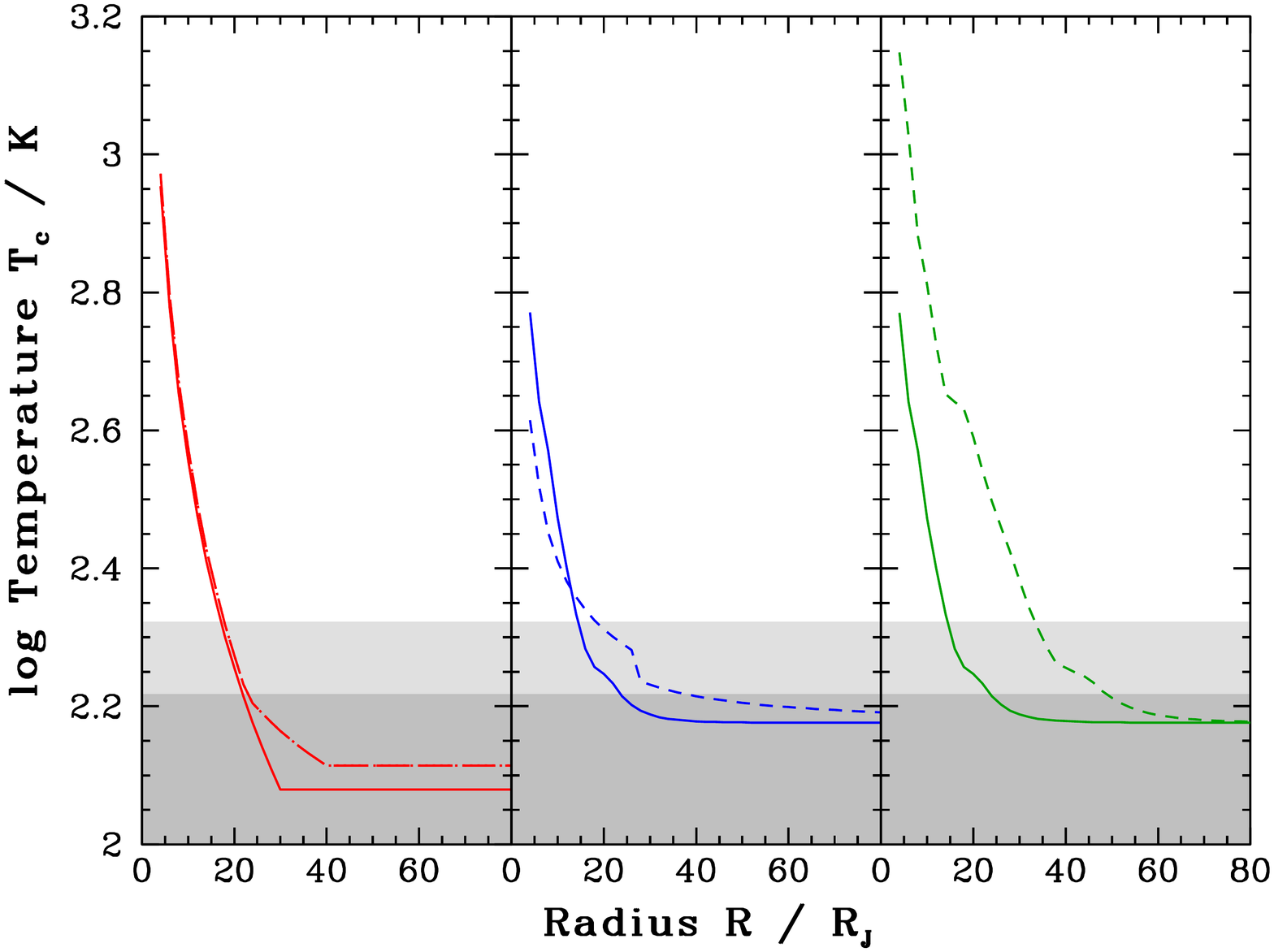}
\caption{\sf The seven subnebula models' radial profiles of surface
  density (top) and midplane temperature (bottom).  The minimum-mass
  models appear at left, the gas-starved models in the center and
  right panels.  Each model's gas and dust surface densities are shown
  by two matching curves, with the gas the larger one.  The three
  minimum-mass models at left are MM96 (solid), MM03 (long-dashed) and
  SEMM (dotted).  Note that the MM03 and SEMM models overlap
  throughout in solid surface density and temperature.  The
  gas-starved models at center differ in opacity and are K0 (solid)
  and K-4 (dashed), while those at right differ in the growth
  timescale and are TG50 (solid) and TG05 (dashed).  Grey shading in
  the surface density panels indicates the range penetrated by X-rays
  (darker) and cosmic rays (lighter).  The shadings in the temperature
  panels indicate the approximate water ice stability range at the
  minimum (darker) and maximum (lighter) pressures found in the
  gas-starved models \citep{2002AJ....124.3404C}.
  \label{fig:sd}}
\end{figure}

The surface density and midplane temperature profiles of the MM96
model are plotted in figure~\ref{fig:sd} along with those of the six
other models described below.

\subsection{Mosqueira \& Estrada (2003) Model --- MM03}

The MM03 model has a more complex surface density profile,
\begin{eqnarray}
\Sigma(r) &=& \Sigma_{\rm in}^0 (R_{\rm in}/r) \qquad {\rm for}\ r \le r_1 \\
&=& a_1 (r_1/r)^{b_1} \qquad {\rm for}\ r_1 \le r \le r_2 \\
&=& \Sigma_{\rm out}^0 (R_{\rm out}/r) \qquad {\rm for}\ r \ge r_2
\end{eqnarray}
where $\Sigma_{\rm in}^0 = 51 \times 10^4$\,g\,cm$^{-2}$,
$\Sigma_{\rm out}^0 = 0.31 \times 10^4$\,g\,cm$^{-2}$, $r_1 = 20 R_J$,
$r_2 = 26 R_J$, $R_{\rm in} = 14 R_J$, $R_{\rm out} = 87 R_J$,
\begin{equation}
a_1 = \Sigma_{\rm in}^0 R_{\rm in}/r_1 = 35.7 \times 10^4\,{\rm g}\,{\rm cm}^{-2},
\end{equation}
and
\begin{equation}
b_1 = \ln[(\Sigma_{\rm in}^0 R_{\rm in} r_2)/(\Sigma_{\rm out}^0
R_{\rm out}
r_1)]/\ln(r_2/r_1) = 13.4871.
\end{equation}
Note that the values of $a_1$ and $b_1$ in Table~2 of
\cite{2003Icar..163..198M} are not exact and that there is a missing
$r_1$ in their eq.~6 for $\Sigma$ in the transition region.

The temperature in Kelvins, based on fitting their Figure~3, is
\begin{eqnarray}
T_c(r) &=& 3750 (R_J/r) \qquad {\rm for}\ r/R_J \le 23 \\
       &=& 655.55 (r/R_J)^{-1/2} + 26.35 \qquad {\rm for}\ 23 \le r/R_J \le 40 \\
       &=& 130 \qquad {\rm for}\ r/R_J \ge 40.
\end{eqnarray}

Like MM96, our MM03 model includes a 1\% mass fraction of solid
material.

\subsection{Solids-Enhanced Minimum Mass Model --- SEMM}

The solids-enhanced minimum mass model preferred by
\cite{2003Icar..163..232M} and \cite{2009euro.book...27E} differs from
the MM03 model in having 90\% of the gas removed within $r_1 = 20 R_J$.
The gas surface density is unchanged outside $r_2 = 26 R_J$, and in
the transition zone between 20 and $26 R_J$ varies smoothly as
$3.57\times 10^4\left(r/r_1\right)^{-4.711}$~g~cm$^{-2}$.
The surface density of the solids is left unchanged throughout.
In this sense, the model is not solids-enhanced but gas-depleted.

\section{IMPROVED GAS-STARVED SUBNEBULA MODEL
  \label{sec:gssn}}

In the gas-starved models, only a fraction of the material needed to
form the satellites orbits the planet at any given instant.  The
subnebula is replenished by the slow inflow of gas and solids after
Jupiter opens a gap in the Solar nebula.  An approximate overall
balance between the growth of new satellites and loss by migrating
into the planet regulates the mass fraction of the satellite system to
$\sim 10^{-4}$ \citep{2006Natur.441..834C}.

Gas-starved models are constructed assuming material from the Solar
nebula falls steadily on the circumplanetary disk
\citep{2002AJ....124.3404C}.  Hydrodynamical calculations treating the
vertical structure show the solar nebula gas approaches the planet and
its disk from above and below \citep{2008ApJ...685.1220M,
  2012ApJ...747...47T, 2012MNRAS.427.2597A}.  The circumplanetary disk
structure is insensitive to the distribution of the injected Solar
nebula gas once a steady-state is reached, depending instead on the
disk's angular momentum balance \citep{2011MNRAS.413.1447M}.  This
contrasts with the minimum-mass models, where the size is fixed by the
angular momentum of the gas at the time the disk is assembled.
Orbital angular momentum is transferred through the gas-starved
subnebula by an unspecified process that yields accretion stresses
equal to a constant, $\alpha$, times the gas pressure
\citep{1973A&A....24..337S}.  The temperature is determined by the
resulting release of gravitational energy, together with the
illumination from Jupiter and from the surrounding Solar nebula,
balanced by radiative losses.

Regarding the circumplanetary disk's size, we can say that the outer
edge lies within 40\% of the planet's Hill radius $r_H$, since at
greater distances the stellar tide is strong and periodic ballistic
orbits cross \citep{2011MNRAS.413.1447M}.  A smaller disk can expand
to $0.4r_H$ under magnetic stresses \citep{2013MNRAS.428.2668L}.  On
the other hand, photoevaporation is capable of truncating
circumplanetary disks to a small fraction of the Hill radius
\citep{2011AJ....142..168M}.  The maximum size of $0.4r_H$ for Jupiter
corresponds to about $300R_J$.

Several further constraints apply to conditions inside Jupiter's disk.
Ganymede's composition requires the water ice sublimation point to lie
inside this moon's orbit when the last generation of satellites form.
Slow growth of the planet before the Solar nebula starts to dissipate
is ruled out because the stellar tides raise a two-armed spiral wave in
hydrodynamical models of an inviscid circumplanetary disk, setting a
floor on the accretion torques that yields a slowest allowed planet
mass doubling time of 5~Myr \citep{2012A&A...548A.116R}.  This
suggests the maximum $\tau_G=10^8$~years used by
\cite{2002AJ....124.3404C} applies only after the Solar nebula starts
to dissipate.  Yet another constraint
comes from \cite{2005A&A...439.1205A} who found that rocky satellites
within 10$R_J$ survive migration if $\alpha > 2\times 10^{-4}$ and
temperatures remain low enough for long enough to form Callisto if
$\alpha < 10^{-3}$.  Note that the heating in their models is
distributed in the disk interior.  Releasing the heat in a
magnetically-active surface layer at lower optical depth yields cooler
midplane temperatures \citep{2011ApJ...732L..30H}.  We therefore do
not attempt to meet the last constraint.

In the gas-starved models of \cite{2002AJ....124.3404C}, the inflowing
material is assumed to be deposited uniformly in the region extending
to distance $r_c$ from the planet, with the total rate of mass inflow
equal to $F_\ast$.  The gas component of the disk spreads viscously,
both onto the planet and out to some assumed outer edge at $r_d$.
Three parameters distinguish the gas-starved models of
\cite{2002AJ....124.3404C}.  These are the stress-to-pressure ratio
$\alpha$; the opacity of the disk to its own radiation, assumed
independent of temperature, density and position; and the rate at
which mass falls on the planet, measured by the planet growth
timescale $\tau_G=M_J/{\dot M_J}$ (where ${\dot M_J} \approx F_\ast$).
We construct versions of these models with three improvements, (1)
making the opacities temperature-dependent, (2) properly treating
optically-thin disk annuli and (3) more accurately computing the
illumination by Jupiter.  Due to the first of these, we replace their
constant opacity parameter $K$ by a dust depletion factor, for which
we use the symbol $f_{\rm opac}=\epsilon/0.01$.  The
temperature-dependent opacities are taken from
\cite{1994ApJ...421..615P}.  The second and third improvements are
made by using midplane temperatures from the analytic vertical
structure model of \cite{1990ApJ...351..632H} for viscous dissipation
and isotropic solar nebula irradiation, with the extension for
irradiation by a central source (i.e.\ Jupiter) by
\cite{2001A&A...379..515M}.

The temperature-dependent opacity for undepleted grain composition is
taken from Figure 6 of \cite{1994ApJ...421..615P}, which includes
contributions from  silicates, troilite, metallic iron, organics, and
water ice. It increases from $0 \,{\rm cm}^2 \,{\rm g}^{-1}$ at $0\,$K
to $6.50 \,{\rm cm}^2 \,{\rm g}^{-1}$ at $174\,$K, and shows multiple
local minima and maxima at higher temperatures, ranging from $1.95
\,{\rm cm}^2 \,{\rm g}^{-1}$ at $700\,$K to $6.28 \,{\rm cm}^2
\,{\rm g}^{-1}$ at $425\,$K.

The product of the disk gas surface density $\Sigma$ and viscosity
$\nu$ is determined by the mass inflow model and independent of the
vertical structure.  So
\begin{equation}
\nu \Sigma = {4 F_* \over 15 \pi} \cases{
{{\textstyle 5} \over {\textstyle 4}} -
\sqrt{{\textstyle r_c} \over {\textstyle r_d}} -
{{\textstyle 1} \over {\textstyle 4}}
\left({\textstyle r} \over {\textstyle r_c}\right)^2 &
for $r < r_c$ , \cr
& \cr
\sqrt{{\textstyle r_c} \over {\textstyle r}} -
\sqrt{{\textstyle r_c} \over {\textstyle r_d}} &
for $r \ge r_c$ ,}
\end{equation}
as in \cite{2002AJ....124.3404C}.  We also follow
\cite{2002AJ....124.3404C} in adopting the $\alpha$ prescription for
the viscosity:
\begin{equation}
\nu = \alpha c_s^2/\Omega,
\end{equation}
where $c_s$ is the midplane sound speed.

We consider heating by viscous dissipation in the disk, incoming
isotropic radiation at the ambient nebular temperature $T_{\rm neb}$,
and incoming radiation from Jupiter.  According to order-of-magnitude
estimates, radial heat advection is unimportant.

The irradiation from our central source, Jupiter, is highly
directional, with the cosine of the characteristic angle (measured
from the inward directed normal to the disk surface) at which the
light enters the disk equal to \citep{1997ApJ...490..368C}
\begin{equation}
\mu_J = {4 \over 3\pi} \left(R_J \over r\right) + 
        \left(H \over r\right) \left({d\ln H \over d\ln r} - 1\right) ,
\end{equation}
and the flux intercepted by a surface element (either top or bottom)
of the disk equal to
\begin{equation}
4\pi H_J(0) = \left(\mu_J \over 2\right) \left(R_J \over r\right)^2
              \sigma_{\rm SB} T_J^4 ,
\label{eq:HJ}
\end{equation}
where $\sigma_{\rm SB}$ is the Stefan-Boltzmann constant.
Equation (\ref{eq:HJ}) was dervied by \cite{1991ApJ...375..740R} in
the limit that $H/r \ll 1$ and $r \gg R_J$ (see also
\citealt{1970PThPh..44.1580K}).
It differs from $4\pi H_J(0) = (dH/dr) (R_J/r)^2 \sigma_{\rm SB}
T_J^4$ used by \cite{2002AJ....124.3404C}.

Energy balance requires the outgoing flux on the disk surface to equal
to the sum of the incoming flux plus the emission from viscous
dissipation.
Thus the {\it net} (outgoing minus incoming) flux on the surface of
the disk is just the emission from viscous dissipation.
If we define the accretion temperature $T_d$ in terms of the net flux,
\begin{equation}
  2 \sigma_{\rm SB} T_d^4 = {9 \over 4} \Omega^2 \nu \Sigma .
\end{equation}

To determine the midplane and surface temperatures, we use the
analytic model of the vertical structure developed by
\cite{1990ApJ...351..632H} for the treatment of viscous dissipation
and isotropic irradiation by the ambient nebula, and we use the
extension of this model by \cite{2001A&A...379..515M} for the
treatment of irradiation by a central source (Jupiter in our case).
We use the simplest form of this model with the following
approximations.  We assume that the different forms of mean opacities
are all equal to the Rosseland mean opacity.  We use the same mean
opacity for the disk's own radiation, the radiation from the ambient
nebula, and the radiation from Jupiter.  We assume that the extinction
is dominated by absorption.  Then the temperature at optical depth
$\tau$ from the surface is given by (see eq.~3.11 of
\cite{1990ApJ...351..632H} and eq.~61 of \cite{2001A&A...379..515M})
\begin{eqnarray}\label{eq:Ttau}
T^4(\tau)
&=& {3 \over 4} \left[\tau \left(1 - {\tau \over 2 \tau_c}\right) +
    {1 \over \sqrt{3}} + {1 \over 3 \tau_c}\right] T_d^4 + T_{\rm neb}^4 \cr
& & \cr
& & + {3 \over 4} \left[\mu_J \left(1 - e^{-\tau/\mu_J}\right) +
    {1 \over \sqrt{3}} + {1 \over 3\mu_J} e^{-\tau/\mu_J}\right]
    \left(4\pi H_J(0) \over \sigma_{\rm SB}\right),
\end{eqnarray}
where $\tau_c$ is the optical depth to the midplane:
\begin{equation}
\tau_c = \kappa \Sigma/2,
\end{equation}
and we use for $\kappa$ the Rosseland mean opacity at the midplane
temperature.
From eq.~\ref{eq:Ttau}, the midplane temperature (at $\tau = \tau_c$)
is given by
\begin{eqnarray}\label{eq:Tctau}
T_c^4
&=& {3 \over 4} \left[{\tau_c \over 2} + {1 \over \sqrt{3}} +
    {1 \over 3 \tau_c}\right] T_d^4 + T_{\rm neb}^4 \cr
& & \cr
& & + {3 \over 4} \left[\mu_J \left(1 - e^{-\tau_c/\mu_J}\right) +
    {1 \over \sqrt{3}} + {1 \over 3\mu_J} e^{-\tau_c/\mu_J}\right]
    \left(\mu_J \over 2\right) \left(R_J \over r\right)^2 T_J^4 ,
\end{eqnarray}
while the surface temperature (at $\tau = 0$) is given by
\begin{equation}
T_s^4 =
{3 \over 4} \left[{1 \over \sqrt{3}} + {1 \over 3 \tau_c}\right] T_d^4 +
T_{\rm neb}^4 + {1 \over 8} \left(R_J \over r\right)^2 T_J^4
\end{equation}
for $\mu_J \ll 1$.
The term in Equation (\ref{eq:Tctau}) with $T_d^4$ due to viscous
heating differs from the expression used by \cite{2002AJ....124.3404C}
in the numerical coefficients and especially in the $1/\tau_c$
dependence in the optically thin limit ($\tau_c \ll 1$).
\cite{2009euro.book...59C} corrected the \cite{2002AJ....124.3404C}
expression for viscous heating for the optically thin regime, but
their expression also differs from Equation (\ref{eq:Tctau}) in the
numerical coefficients.

If $\tau_c \ll \mu_J \ll 1$,
\begin{equation}
T_c^4 \approx
T_s^4 \approx {1 \over 4\tau_c} T_d^4 + T_{\rm neb}^4 +
              {1 \over 8} \left(R_J \over r\right)^2 T_J^4 .
\end{equation}
If $\tau_c \ll 1$ but $\tau_c \not\ll \mu_J$, the last term of
eq.~\ref{eq:Tctau} cannot be reduced to $(1/8) (R_J/r)^2 T_J^4$.
If $\tau_c \gg 1$,
\begin{equation}
T_c^4 \approx {3 \tau_c \over 8} T_d^4 + T_{\rm neb}^4 +
              {\sqrt{3} \over 4} \left(\mu_J \over 2\right)
              \left(R_J \over r\right)^2 T_J^4 ,
\end{equation}
and
\begin{equation}
T_s^4 \approx {\sqrt{3} \over 4} T_d^4 + T_{\rm neb}^4 +
              {1 \over 8} \left(R_J \over r\right)^2 T_J^4 .
\end{equation}

We consider four gas-starved models, two of which are similar to the
high ($K=1 \,{\rm cm}^2 \,{\rm g}^{-1}$) and low ($K=10^{-4} \,{\rm
  cm}^2 \,{\rm g}^{-1}$) opacity models considered by
\cite{2002AJ....124.3404C}.  For the $f_{\rm opac}=1$ model, which is
optically thick out to $\sim 60 R_J$, we only need to decrease
$\tau_G$ slightly to produce surface density and midplane temperature
profiles that are similar to those shown in Figure~6 of
\cite{2002AJ....124.3404C}.  For the $f_{\rm opac}=10^{-4}$ model,
almost the entire subnebula is optically thin, and we obtain
significantly higher midplane temperatures with our improved vertical
structure model (which radiates away the accretion power inefficiently
in the optically thin regime), if we take the parameters from Figure 5
of \cite{2002AJ....124.3404C}.  To place the ice sublimation front
near Ganymede's orbit, we therefore use a lower stress parameter
$\alpha$ and a longer growth timescale $\tau_G$.  Since the improved
and \citet{2002AJ....124.3404C} models have similar surface density
and midplane temperature profiles after suitably adjusting $\alpha$
and $\tau_G$, the models are expected to have similar ionization
states.  However the satellites' orbital migration as they interact
with the disk can be very different in the improved models, due to
sharp jumps in the local power-law indices of the surface density and
midplane temperature profiles resulting from the temperature-dependent
opacity (Li \& Lee, in preparation).

In addition, we consider a scenario where the total rate of mass
inflow to the disk is initially nearly constant at $F_\ast(0)$ and
then decays exponentially with time, i.e., $F_\ast(t) = F_\ast(0)
\exp(-t/\tau_{\rm in})$, due to the dispersal of the Solar nebula
\citep{2006Natur.441..834C, 2008Icar..198..163B, 2012ApJ...753...60O}.
During the exponential decay, the total mass delivered to the disk
after time $t$ is $F_\ast(t) \tau_{\rm in}$.  The last generation of
satellites has total mass $M_T$ and forms after time $t_s$, where
$F_\ast(t_s) \tau_{\rm in} = M_T/\Phi$ and the overall solids-to-gas
ratio in the inflow $\Phi \approx 0.01$.  Note that $\Phi$ is a
surface-integrated measure of the infalling material and need not
match $\phi(r)$, the disk's internal radial profile.  So $\tau_G(t_s)
= M_J/F_\ast(t_s) = (M_J\Phi/M_T) \tau_{\rm in}$.  For the jovian
satellites, if $\tau_{\rm in} \approx 1\,$Myr, $\tau_G(t_s) \approx
50\,$Myr.  Thus we consider two models with $\tau_G (0) = 5\,$Myr and
$\tau_G (t_s) = 50\,$Myr.

The four specific gas-starved models we construct have the parameters
listed in Table~\ref{tab:gs}, as well as $r_c = 30 R_J$, $r_d = 150
R_J$, $T_{\rm neb} = 150\,{\rm K}$, and $T_J = 500\,{\rm K}$, and
their surface density and midplane temperature profiles are shown in
Figure~\ref{fig:sd}.  Treating the opacities' temperature dependence
leads to several new features.  For example, the steep $T_c$ drop near
$27R_J$ in the K-4 model results from a sharp decline in the opacity
at 174~K where water ice sublimates.  The model is marginally
optically-thin at this location.  Temperatures are significantly
higher in the $\tau_G (0) = 5\,$Myr model, which is allowed as there
are no compositional constraints on the earlier generations of
satellites lost by migration into the planets.

\begin{table}[tb]
  \caption{\sf The four gas-starved subnebula models.
    \vspace*{3mm}
    \label{tab:gs}}
\begin{center}
\begin{tabular}{llrl}
\hline
Name  & $\alpha$ & $\tau_G$ (Myr) & $f_{\rm opac}$ \\
\hline
\hline
K0    & 0.005   & 70             & 1         \\
%K$-$2& 0.0025  & 8              & 10$^{-2}$ \\
K$-$4 & 0.0009  & 20             & 10$^{-4}$ \\
TG05  & 0.001   &  5             & 0.1 \\
TG50  & 0.001   & 50             & 0.1 \\
\hline
\end{tabular}  
\end{center}
\end{table}  

%%%%%%%%%%%%%%%%%%%%%%%%%%%%%%%%%%%%%%%%%%%%%%%%%%%%%%%%%%%%%%%%%%%%%%%%%%%%%%%
\section{IONIZATION STATE\label{sec:ionization}}

\subsection{Chemical Network\label{sec:network}}

The ionization state is calculated by integrating a chemical network
treating in simplified form the most important gas-phase pathways:
molecular ionization, dissociative molecular recombination, charge
transfer to metal atoms, and radiative recombination of metal ions.
The closed and balanced set of reactions related to the representative
molecular ion HCO$^+$ is
\begin{eqnarray}
  {\rm H}_2   + X         &\rightarrow& {\rm H}_2^+ + e^-\\
  {\rm H}_2^+ + {\rm H}_2 &\rightarrow& {\rm H}_3^+ + {\rm H}\\
  {\rm H}_3^+ + {\rm CO}  &\rightarrow& {\rm HCO}^+ + {\rm H}_2\\
  2 {\rm H}   + G         &\rightarrow& {\rm H}_2   + G\\
  {\rm HCO}^+ + e^-       &\rightarrow& {\rm CO}   + {\rm H} \label{eq:dr}
\end{eqnarray}
Here every species (except the X-rays, $X$ and grains, $G$) is created
in at least one reaction and destroyed in at least one other.  Over
the whole set, no species is produced or consumed on balance.

The subset producing the ions and electrons, eqs. 26-29, boils down to
\begin{equation}
    2{\rm H}_2 + 2X + 2{\rm CO} \rightarrow {\rm H}_2 + 2{\rm
      HCO}^+ + 2 e^- \label{eq:molecularion}.
\end{equation}
That is, each X-ray striking a hydrogen molecule yields one ion and
one electron.  We follow \citet{2006A&A...445..205I} and others in
approximating eqs.~\ref{eq:molecularion} and~\ref{eq:dr} by
\begin{eqnarray}
  {\rm H}_2 + X &\rightarrow& {\rm HCO}^+ + e^-\label{eq:mi}\\
  {\rm HCO}^+ + e^- &\rightarrow& {\rm H}_2,\label{eq:dr2}
\end{eqnarray}
neglecting the fact that the molecular ion holds just one hydrogen
atom.  This is fine since the HCO$^+$ is orders of magnitude less
abundant than the H$_2$ and forming the ions leaves the H$_2$ density
basically unchanged.  Similarly, we don't follow CO destruction and
reformation since the ion is so much less abundant than the molecule.
To eqs.~\ref{eq:mi} and~\ref{eq:dr2} we add the charge transfer and
radiative recombination reactions involving the representative metal
magnesium:
\begin{eqnarray}
  {\rm HCO}^+ + {\rm Mg} &\rightarrow& {\rm HCO} + {\rm Mg}^+\label{eq:ct}\\
  {\rm Mg}^+  + e^- &\rightarrow& {\rm Mg} + h\nu.
\end{eqnarray}
The product radical in eq.~\ref{eq:ct} readily breaks apart into H and
CO.  Simplifying by again taking into account the large abundances of
H$_2$ relative to H, and CO relative to HCO$^+$, we arrive at the
reduced network
\begin{eqnarray}
  {\rm H}_2 + X &\rightarrow& {\rm HCO}^+ + e^-\\
  {\rm HCO}^+ + e^- &\rightarrow& {\rm H}_2\label{eq:molrec}\\
  {\rm HCO}^+ + {\rm Mg} &\rightarrow& {\rm H}_2 + {\rm Mg}^+\\
  {\rm Mg}^+  + e^- &\rightarrow& {\rm Mg} + h\nu.
\end{eqnarray}
Whether the carbon is oxidized or reduced makes little difference here
because (1) the rate coefficient for pathway~\ref{eq:molrec} at
$3\times 10^{-6}/\sqrt{T}$~cm$^3$~s$^{-1}$ \citep{2006A&A...445..205I}
is similar to that for the corresponding methane ion, $10^{-6}$
\citep{1996Icar..123..404T}, and (2) anyway gas-phase recombination
proves less important than the grain surface pathway, near the dead
zone boundary in our cases with dust.

Also treated in the network are grain charging and discharging
\citep{2006A&A...445..205I} through collisions with ions and
electrons, and charge exchange in grain-grain collisions.  Grain
charges from $-2$ to $+2$ are considered.  Additionally the metal
atoms are allowed to thermally adsorb on and desorb from the grains.
The reactions and their rate coefficients are described by
\cite{2006A&A...445..205I}, with the electron sticking probabilities
revised to include the grain charge following
\cite{2011ApJ...739...50B}.  The magnesium locked up inside grains is
assumed to be 99\% of the Solar abundance of $3.7\times 10^{-5}$ per
hydrogen atom, with the remaining 1\% available to participate in the
recombination network, either in the gas phase or adsorbed on grain
surfaces.  The gas-phase magnesium abundance had little effect on the
magnetic activity above a threshold level of $10^{-6}$ times Solar, in
protostellar disk models by \citet{2007ApJ...659..729T}.  We solve the
kinetic equations describing the reaction network using a
semi-implicit extrapolation method.  While bringing the network to
equilibrium we record the recombination time $t_{\rm rec}$ needed to
reach an electron fraction within 1\% of the equilibrium value.

We include monodisperse grains $a=0.1$~$\mu$m in radius with internal
density $\rho_d=2$~g~cm$^{-3}$.  This yields a geometric cross-section
per unit dust mass similar to that of the size distribution used to
compute the opacities by \cite{1985Icar...64..471P} and
\cite{1994ApJ...421..615P}.  Furthermore the same dust-to-gas ratios
are used for the opacities and the grain surface recombination in the
dusty versions of our gas-starved models.

\subsection{Ionization Processes}

The chemical reaction network is driven by the ionization from
interstellar cosmic rays, radioisotope decay and protosolar X-rays.
The cosmic rays yield an ionization rate $10^{-17}$~s$^{-1}$ well
outside the Solar nebula.  They strike our material isotropically over
the upper hemisphere and their secondary particles are absorbed over a
column 96~g cm$^{-2}$ following \cite{1981PASJ...33..617U} and
\cite{2009ApJ...690...69U}.

We consider two radioisotope ionization scenarios.  Long-lived
isotopes such as potassium-40 yield an ionization rate $6.9\times
10^{-23}(\epsilon/0.01)$~s$^{-1}$, while short-lived isotopes such as
aluminum-26 if present yield a much higher rate, $3.7\times
10^{-19}(\epsilon/0.01)$, where $\epsilon$ is the dust-to-gas mass
ratio \citep{1992Icar...97..130S, 1996Icar..123..404T,
  2009ApJ...690...69U, 2009Icar..204..658C}.

The Solar nebula for most of its lifetime blocks direct sightlines so
that the protosolar X-rays reach the planet's vicinity entirely
through scattering.  Jupiter and its disk at one time lay in a gap in
the Solar nebula \citep{1993prpl.conf..749L} and the geometry of the
gap surely influenced the flux of X-rays reaching the planet.
Furthermore, toward the end of the Solar nebula's evolution the gas
interior to Jupiter's orbit cleared first, judging from the central
holes observed in the so-called transitional systems found among
protostellar disks today \citep{2005ApJ...630L.185C,
  2010ApJ...708.1107M, 2011ApJ...732...42A}.  Jupiter and surrounding
material were then directly exposed to protosolar X-rays.  However
lacking detailed information about the X-ray transfer in either of
these geometries, we use the ionization rates vs.\ column in the Solar
nebula derived from Monte Carlo transfer calculations by
\citet{1999ApJ...518..848I}, taking the case with the 5~keV thermal
spectrum from their figure~3 and scaling the luminosity to $2\times
10^{30}$~erg~s$^{-1}$, the median observed in young Solar-mass stars
in the Orion Nebula Cluster \citep{2000AJ....120.1426G}.  The
scattered X-rays are absorbed in a column of about 8~g~cm$^{-2}$.

\begin{figure}[htb!]
  \epsscale{0.6} \plotone{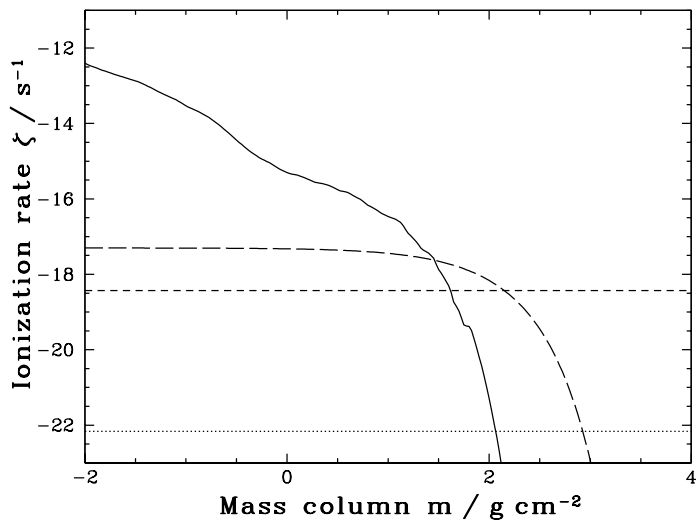}
  \caption{\sf Ionization rates per H nucleus vs.\ the column of
    overlying material, resulting from X-rays (solid curve), cosmic
    rays (long-dashed curve) and short- and long-lived radionuclides
    (dashed and dotted lines).  Only the X-rays and cosmic rays
    arriving from above are included.  The X-ray ionization rate is
    from Monte Carlo calculations of the transfer through the Solar
    nebula of photons with a 5~keV thermal spectrum
    \citep{1999ApJ...518..848I}.  The X-rays are extrapolated past
    80~g~cm$^{-2}$ using the $e$-folding depth of their scattered
    component, 8~g~cm$^{-2}$ \citep{2008ApJ...679L.131T} while the
    cosmic rays' $e$-folding depth is 96~g~cm$^{-2}$
    \citep{1981PASJ...33..617U, 2009ApJ...690...69U}.  The
    radionuclide abundances correspond to the interstellar dust-to-gas
    mass ratio~$\epsilon=0.01$.  \label{fig:ioniz}}
\end{figure}

The ionization rate contributions from all the non-thermal processes
are shown as functions of the mass column in figure~\ref{fig:ioniz}.

Finally we treat the thermal ionization of the
low-ionization-potential element potassium, which becomes important at
the temperatures above $1\,000$~K reached inside Io's orbit
\citep{1996Icar..123..404T}.  We solve the Saha equation using
potassium's ionization energy of 4.3407~eV and assuming 99\% of the
Solar potassium abundance is locked up inside the grains, with the
remaining 1\% in the gas and available for collisional ionization.
The steep temperature dependence of the thermal ionization means the
dead zone boundary is insensitive to this choice.

%%%%%%%%%%%%%%%%%%%%%%%%%%%%%%%%%%%%%%%%%%%%%%%%%%%%%%%%%%%%%%%%%%%%%%%%%%%%%%%
\section{MAGNETO-ROTATIONAL TURBULENCE\label{sec:mri}}

Analytic and numerical results indicate that the criterion for
magneto-rotational instability to drive turbulence is
\begin{equation}\label{eq:elsasser}
  \Lambda \equiv {v_{Az}^2\over\eta\Omega} > 1,
\end{equation}
where the dimensionless Elsasser number $\Lambda$ depends on the
Alfv\'{e}n speed $v_{Az}$ for the vertical component of the magnetic
field, along with the magnetic diffusivity $\eta$ and orbital
frequency $\Omega$.  This means the instability must grow faster than
the magnetic fields can diffuse across its fastest-growing wavelength
\citep{1996ApJ...457..798J, 1999ApJ...515..776S, 2001ApJ...561L.179S,
  2002ApJ...577..534S, 2007ApJ...659..729T}.  In the diffusivity we
include the contributions from the induction equation's Ohmic and
ambipolar terms, added in quadrature.  A further requirement for the
instability to grow near its top rate is that the background toroidal
magnetic fields have a pressure less than the gas pressure
\citep{2000ApJ...540..372K}.

We compute the diffusivity $\eta$ including the current densities from
all the charged species in the chemical network described in
section~\ref{sec:ionization}, following eqs.~21 to~31 of
\cite{2007Ap&SS.311...35W}.  Both the Ohmic and ambipolar terms in the
induction equation are included.  Ohmic diffusion occurs at densities
high enough for the main charged species to couple to the neutrals
through collisions, while ambipolar drift is important when densities
are low enough and collisions rare enough that the neutrals slip
through the plasma which remains tied to the magnetic fields by
Lorentz forces.  At intermediate densities a third non-ideal effect,
the Hall term, is important \citep{1999MNRAS.303..239W}.  We neglect
the Hall term because it affects the turbulence threshold and
saturation level only slightly when comparable to the Ohmic term
\citep{2002ApJ...577..534S}.  However, dramatic effects appear in
unstratified non-linear calculations when the Hall term dominates
\citep{2013MNRAS.434.2295K}.  Stratified calculations are urgently
needed.

The maximum possible accretion stress depends on the Elsasser number
$\Lambda$.  When the diffusivity is dominated by the ambipolar term,
the stress can reach about 1\% of the gas pressure if $\Lambda\approx
1$, and 10\% if $\Lambda\approx 10$, according to 3-D unstratified
shearing-box MHD results \citep{2011ApJ...736..144B}.  The higher of
these stress levels can occur starting one decade above our magnetic
activity threshold, or one contour level in the plots in
section~\ref{sec:deadzones} below.

\subsection{Magnetic Fields}

The magneto-rotational instability grows from initially-weak magnetic
fields into long-lived turbulence in both local and global MHD
calculations \citep{2000ApJ...534..398M, 2006A&A...457..343F}.  Over a
wide range of seed field strengths, the pressure in the fields'
vertical component saturates between $10^{-4}$ and $10^{-2}$ times the
midplane gas pressure \citep{2000ApJ...534..398M, 2006A&A...457..343F,
  2010ApJ...708.1716S, 2010MNRAS.409.1297F, 2011ApJ...742...65O}.
Owing to the fields' buoyancy, the magnetic pressure declines more
slowly with height than the gas pressure.  We therefore compute the
Elsasser number at each point assuming that the pressure in the
vertical component of the magnetic field is simply 0.1\% of the
midplane gas pressure, independent of height.  Note that this measures
not just the net vertical or seed magnetic field delivered with the
gas arriving from the Solar nebula, but the overall vertical field
including the part generated locally in the turbulence.  We seek
places in the circumjovian disk where turbulence can be sustained.

Choosing $10^{-3}$ for the midplane pressure ratio means the magnetic
field's vertical component has pressure greater than the gas above
3.7~density scale heights.  In saturated MRI turbulence, the toroidal
magnetic field has a pressure at least ten times the vertical
component \citep{2000ApJ...534..398M}, giving a total magnetic
pressure exceeding the gas pressure above about $3H$.  The MRI's
linear growth rate is reduced at low plasma beta
\citep{2000ApJ...540..372K} so turbulence would be increasingly weaker
above this height.  However there would be less weakening if we
included (1) the field strength's fall-off above a few scale-heights,
and (2) the gas pressure profile's extended tail resulting from
magnetic support.  Both these effects increase the plasma beta over
our simple picture, and both are observed in the 3-D numerical
calculations cited above.

Jupiter's magnetic field can safely be neglected since it is weaker
than the MRI-generated fields in all our disks.  This is true if the
planet's field strength is 10~Gauss at its surface, located at 2$R_J$,
and falls off like a dipole in proportion to the inverse cube of the
radius.

%%%%%%%%%%%%%%%%%%%%%%%%%%%%%%%%%%%%%%%%%%%%%%%%%%%%%%%%%%%%%%%%%%%%%%%%%%%%%%%
\section{DEAD ZONES\label{sec:deadzones}}

\subsection{Tests}

As a test, we begin by replicating the \cite{1996Icar..123..404T}
findings under ionization by short-lived radioisotopes and cosmic
rays.  Their gas-phase reaction network differs from ours in lacking
charge transfer to metal atoms, including instead charge transfer to
methane, ammonia and water molecules.  They consider grains 1~cm in
radius, which contribute negligibly to the overall recombination
cross-section.  Our network can be made very like theirs by removing
the dust and replacing the magnesium with a generic molecule having
abundance $10^{-3}$ per hydrogen atom and recombination rate
coefficient $10^{-6}$~cm$^3$~s$^{-1}$.  The resulting ionization
fractions in the MM96 model disk closely follow
\cite{1996Icar..123..404T} figures~4 (b), (c) and (f).

Next we restore the gas-phase reaction network described in
section~\ref{sec:ionization}, including the metal ions which are
long-lived due to their much smaller recombination coefficient.  The
corresponding ionization fractions in the disk interior are about
three orders of magnitude greater.  This is consistent with the
picture in protostellar disks, where the metal atoms play a
significant role when the dust abundance is low
\citep{2002MNRAS.329...18F, 2013ApJ...765..114D}.

\subsection{Fiducial Models}

\begin{figure}[tb!]
\epsscale{.5}
\plotone{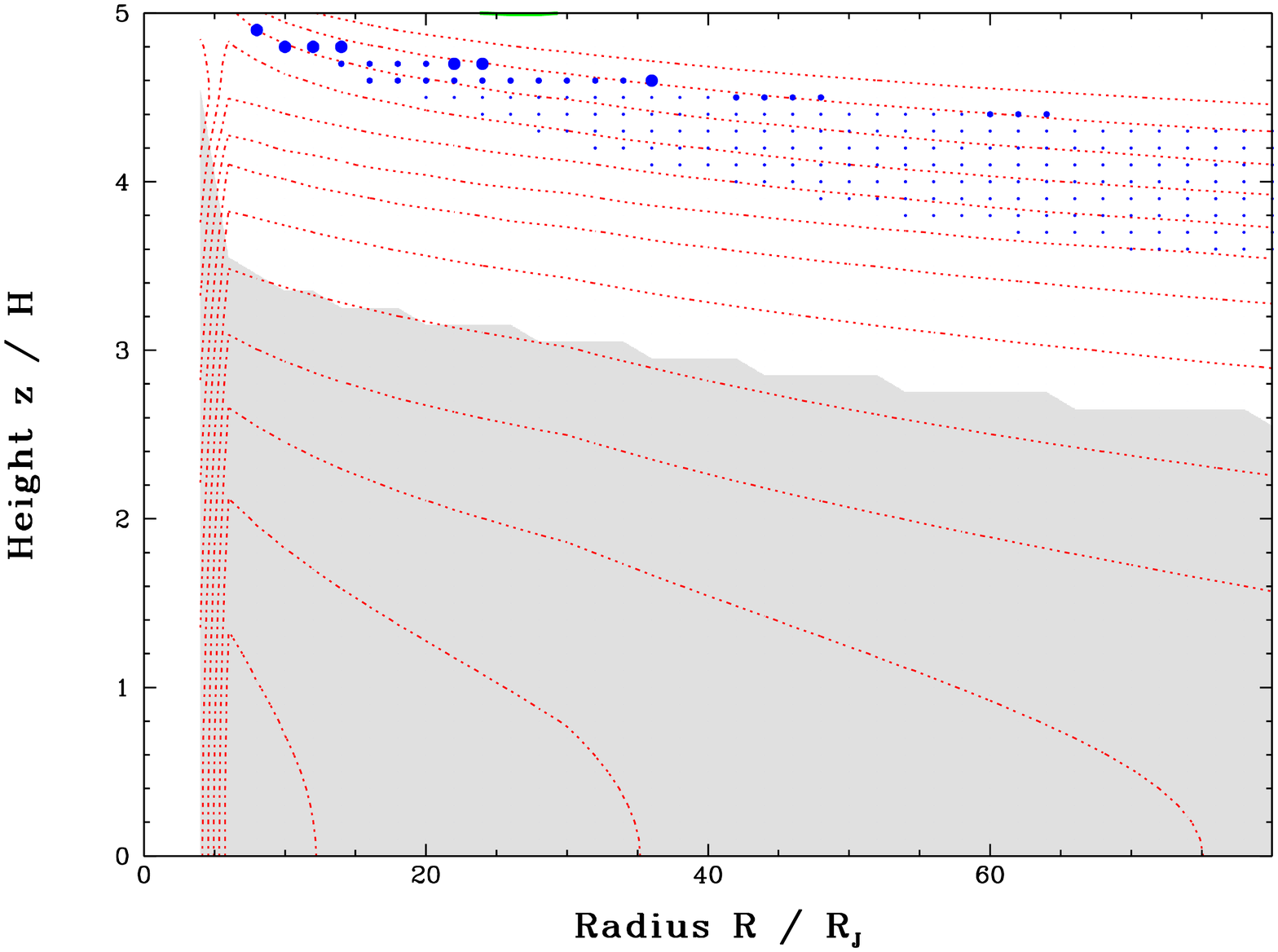}\\
\plotone{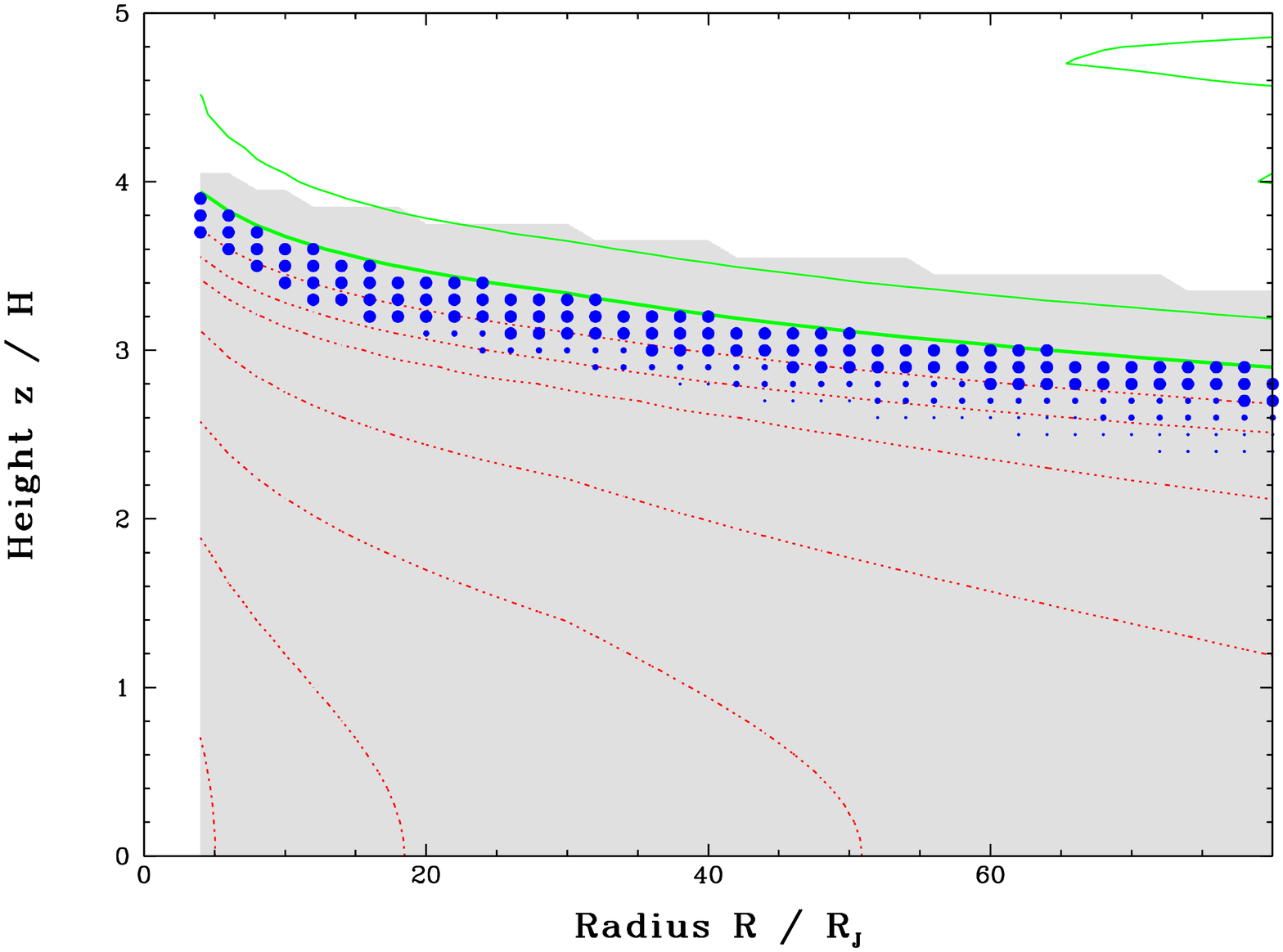}
\caption{\sf The minimum-mass MM96 model has a substantial dead zone
  even in the most favorable ionization scenario with X-rays, cosmic
  rays and short-lived radionuclides.  The dusty case is above, the
  dust-free case below.  In the grey shaded region the Ohmic
  diffusivity is greater than the ambipolar diffusivity.  The contours
  show Elsasser numbers computed from the quadrature sum of the two.
  The Elsasser number is unity on the heavy green contour.  Other
  contours are spaced by an order of magnitude, with the solid green
  ones on the MRI-unstable side and the dashed red ones on the
  MRI-stable side.  The uppermost dashed contour in the dusty case is
  for Elsasser number~0.1.  Blue dots mark where turbulent mixing is
  capable of affecting the dead zone's diffusivity.  At the smallest
  blue dots, the overlying column of free electrons is sufficient to
  lift the diffusivity above the threshold for turbulence if
  instantaneously well-mixed.  At the medium blue dots, the
  recombination time is also at least 10\% of the turbulent mixing
  time.  At the largest blue dots, recombination is slower than
  mixing.  \label{fig:deadupclose2}}
\end{figure}

We then consider the most favorable situation for the MRI, with
ionization by X-rays, cosmic rays and short-lived radionuclides,
looking at one fiducial model each from the minimum-mass and
gas-starved classes.  The minimum-mass MM96 model is shown in
figure~\ref{fig:deadupclose2}.  With recombination on grains (top
panel), the dead zone extends to five scale heights and above, where
the low gas densities mean the boundary is set by ambipolar diffusion.
Combined with the low plasma beta above $5H$, this means MRI
turbulence is weak or absent throughout.  In contrast, without dust
(bottom panel) the dead zone extends only up to $3H$, leaving a small
fraction of the mass column magnetically active.  The active layer's
lower boundary is set by Ohmic diffusion over its whole length.

The two recombination scenarios shown in figure~\ref{fig:deadupclose2}
yield similar outcomes when applied to the steeper surface density
profile $\Sigma=10^7(R_J/R)^{1.3}$~g~cm$^{-2}$ originally suggested by
\cite{1982Icar...52...14L}.

\begin{figure}[tb!]
\epsscale{.5}
\plotone{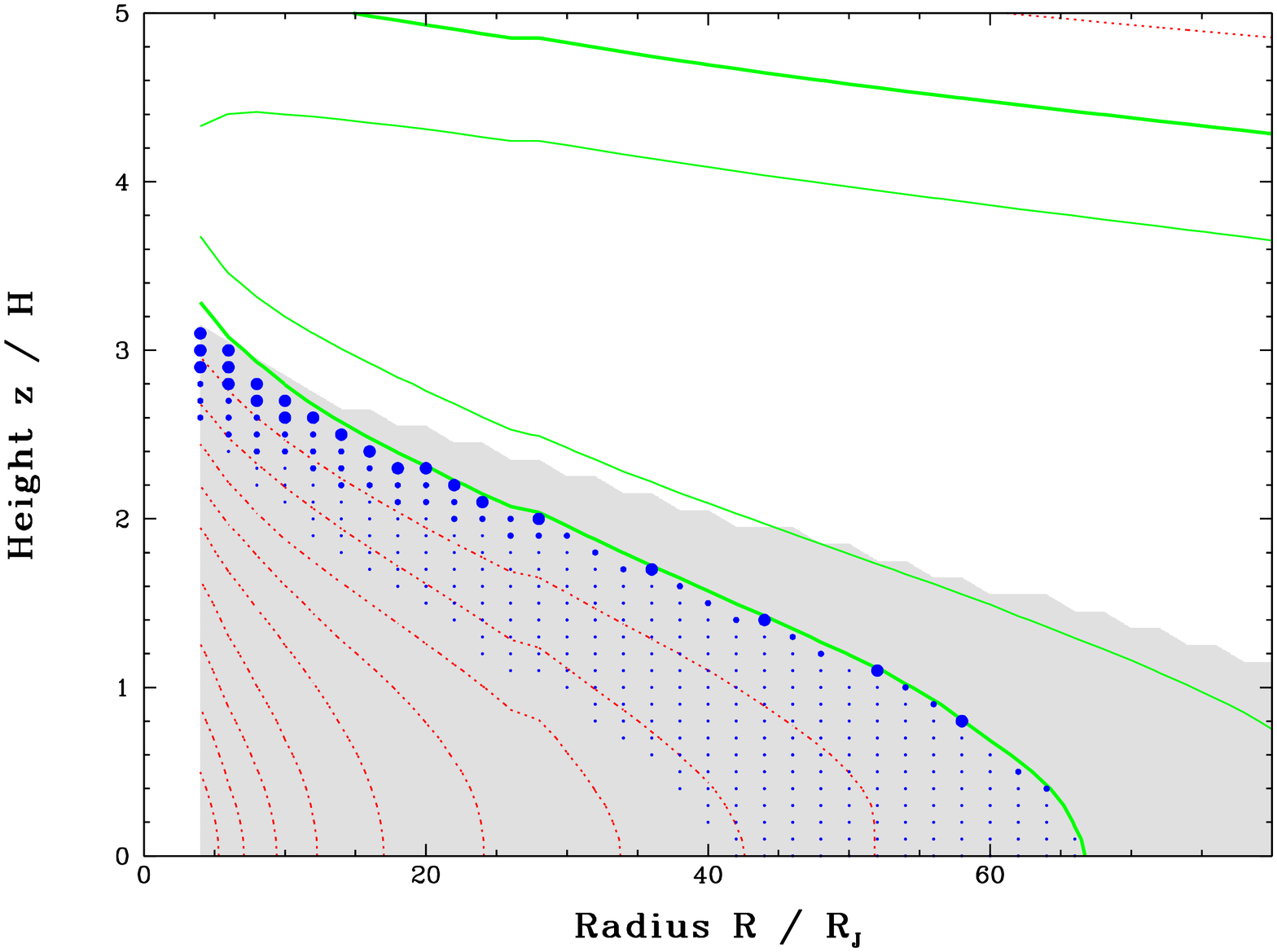}\\
\plotone{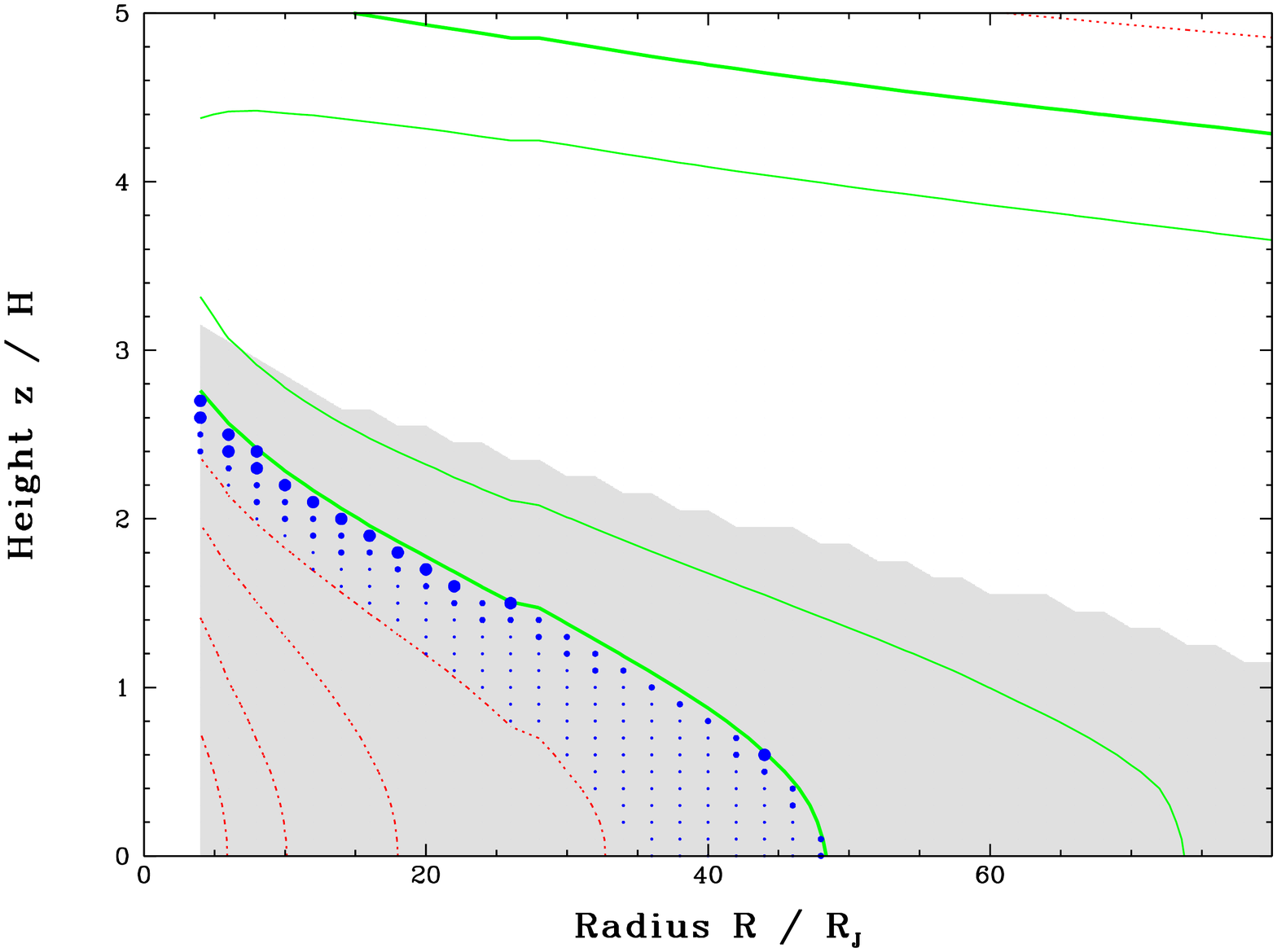}
\caption{\sf Magnetically-active layer and dead zone in the
  gas-starved K-4 model with ionization by X-rays, cosmic rays and
  short-lived radionuclides.  Recombination on grains is included in
  the top panel only.  Contours, shading and symbols are as in
  figure~\ref{fig:deadupclose2}.  \label{fig:deadupclose}}
\end{figure}

As a fiducial gas-starved disk we choose the K-4 model shown in
figure~\ref{fig:deadupclose}.  In contrast to the fiducial
minimum-mass model, the disk here resembles the Solar nebula in having
a substantial magnetically-active surface layer overlying an interior
dead zone.  Over almost the whole radial extent of our calculation,
the boundary between the two layers lies below the height of $3H$
where the plasma beta falls to unity.  MRI can therefore grow at near
its maximum rate.  Over a similar radial range, the Ohmic term
dominates the magnetic diffusivity at the boundary.  Basically the
whole mass column is active beyond $67R_J$ with dust, or $48R_J$
without.

\subsection{Turbulent Mixing}

We also consider turbulent mixing, which alters the resistivity if the
mixing is faster than the chemical reactions.  In geometrically-thin
accretion disks, vertical gradients are generally steeper than radial
gradients, so the greatest effects come from mixing in the vertical
direction.  Representing the mixing as a diffusion process, we can
write the mixing time as the ratio of the squared density scale height
to the diffusion coefficient.  The diffusion coefficient in MRI
turbulence is approximately the mean squared velocity dispersion
divided by the shear rate $\frac{3}{2}\Omega$, and the velocities are
roughly equal to the Alfv\'{e}n speeds.  The vertical mixing timescale
is therefore
\begin{equation}
{t_{\rm mix}} = {3\over 8\pi} \left({2c^2\over v_{Az}^2}\right)
  {2\pi\over\Omega}.
\end{equation}
Using the result that MRI turbulence leads to tangled magnetic fields
in which the vertical component contributes around 10\% of the
magnetic pressure \citep{2000ApJ...534..398M}, we can say that the
number of orbits needed to mix through one scale height is about equal
to the plasma beta parameter $\beta=2c^2/v_A^2$ for the total magnetic
field.

This result lets us estimate the importance of mixing relative to
recombination in the circumjovian disk, compared with the nearby Solar
nebula.  The ratio of the turbulent mixing timescale
$\sim\beta/\Omega$ to the recombination timescale $\sim 1/\rho^2$ is
greater in the small disk in proportion to $\rho^2/\Omega \sim
(\Sigma/H)^2/\Omega \sim (1/10^{-3})^2/10^3 = 1000$.  That is, mixing
is less effective in the circumjovian disk.  Here we obtained a lower
bound by using a surface density $\sim 100$~g~cm$^{-2}$ from the
gas-starved disks, where recombination is slowest among the
circumplanetary models.  Also, we took similar chemical compositions
so that recombination rates are simply proportional to density
squared, and we assumed that the MRI turbulence saturates at similar
plasma beta values in the two situations, giving comparable mixing
timescales when measured in local orbits.

For a more thorough evaluation taking into account the different
chemical composition, consider a blob of gas high in the disk
atmosphere where the ionization is substantial.  The question is
whether, as the blob is carried downward in the turbulence, the
electron fraction remains greater than the ambient equilibrium value.
To answer, we approximate the reacting flow by moving the blob
instantaneously to the interior point of interest.  If recombination
brings the ionization fraction to equilibrium at the new location over
a time at least comparable to the turbulent mixing time, then the
mixing can change the ionization state.  Sometimes we can avoid even
the complication of integrating the chemical network.  If mixing the
overlying column thoroughly and instantaneously would yield a magnetic
diffusivity too high for MRI turbulence, the point of interest will
remain dead.  We focus on the electrons' mixing, since the transport
shifts only the turbulent layer's bottom boundary, which typically is
set by the Ohmic diffusivity, and the Ohmic term is controlled by the
electron fraction.  The procedure in detail is as follows:
\begin{enumerate}
\item Determine the local chemical equilibrium ionization states at
  all heights in some annulus of the model disk.
\item Consider a height $z_1$ where the electron fraction is below the
  MRI threshold or critical value $x_c(z_1)$.  In local chemical
  equilibrium, the point $z_1$ is magnetically dead.  Integrate the
  columns of free electrons and of neutrals above $z_1$ to find the
  lowest height $z_2>z_1$ such that the mean electron fraction between
  $z_1$ and $z_2$ exceeds the critical value $x_c$.  If no $z_2$
  within the model disk yields a mean electron fraction exceeding the
  critical value, then even instantaneous mixing would not be
  effective, and point $z_1$ will remain dead.
\item Take the abundances of all species $j$ from $z_2$ and insert
  them at $z_1$.  That is, the number density $n_{1j}$ is equal to
  $n_{2j}(n_{1n}/n_{2n})$, where the subscripts $n$ indicate the
  background neutral component.  Find the recombination time by
  integrating the reaction network at $z_1$ till the electron fraction
  drops below the critical value $x_c$.
\item The mixing time between the two heights is $(z_2-z_1)^2/D$,
  where $D=v_{Az}^2/\Omega$ is the turbulent diffusion coefficient
  measured at $z_1$.  The diffusion coefficient generally increases
  with height, so the value at the bottom determines the mixing
  timescale.
\item Mixing can alter the ionization fraction at $z_1$ if
  recombination is not much faster than mixing.  We require
  $t_{rec}>0.1 t_{mix}$.
\end{enumerate}
Similar results come from extending step~2 to consider all heights,
and not just the lowest point $z_2$ having $x_{e,avg}>x_{e,c}$.  This
is because the biggest ratio of recombination to mixing time typically
is found at or just above $z_2$.

\subsection{Parameter Survey}

Putting together the turbulence criterion and the mixing timescale, we
infer that the magnetic fields drive turbulence where either (1) the
equilibrium ionization is strong enough by eq.~\ref{eq:elsasser}, or
(2) MRI occurs in the overlying layers in the equilibrium ionization
state, and the resulting mixing is fast enough to raise local
ionization levels to the threshold.  Below we calculate the extent of
the turbulence for several models of the Jovian subnebula.

\begin{figure}[htb!]
\epsscale{1}
\plotone{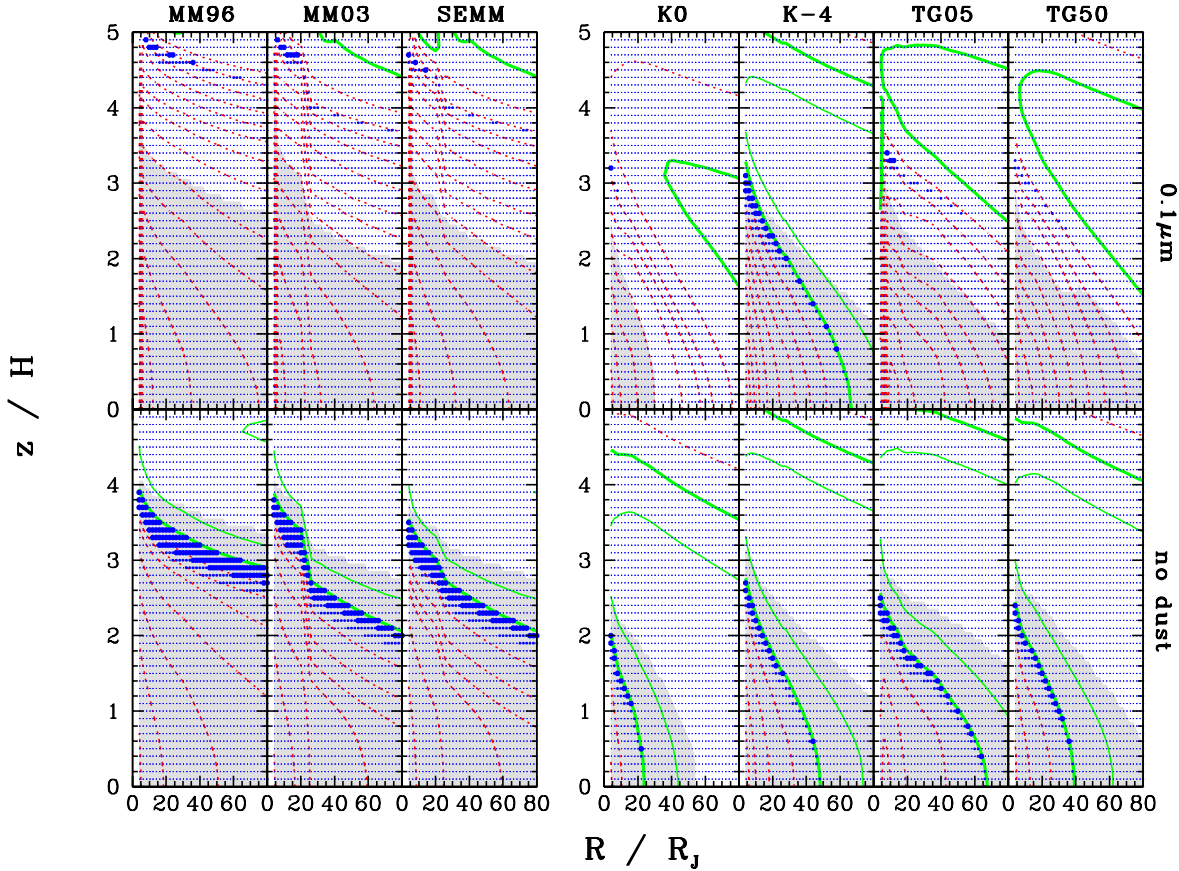}
\caption{\sf Magnetically-active layers and dead zones under
  ionization by X-rays, cosmic rays and short-lived radionuclides, in
  the three minimum-mass (left side) and four gas-starved Jovian
  subnebula models (right side), with and without recombination on
  dust (top and bottom rows, respectively).  Symbols are as in
  figure~\ref{fig:deadupclose2}.  \label{fig:deadxrcrsr}}
\end{figure}

\begin{figure}[htb!]
\epsscale{1}
\plotone{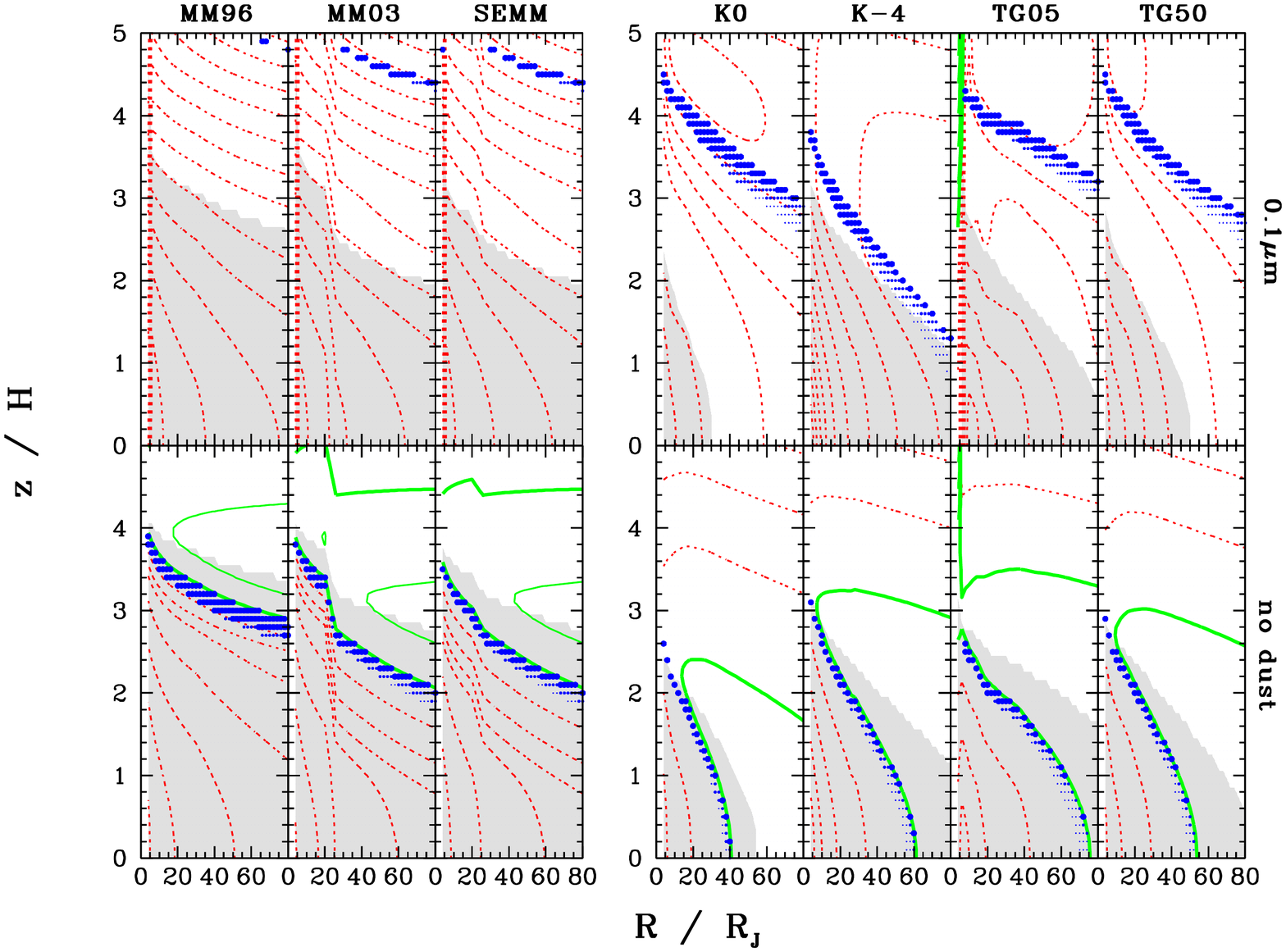}
\caption{\sf As figure~\ref{fig:deadxrcrsr} except that the X-ray
  ionization is omitted.  \label{fig:deadcrsr}}
\end{figure}

\begin{figure}[htb!]
\epsscale{1}
\plotone{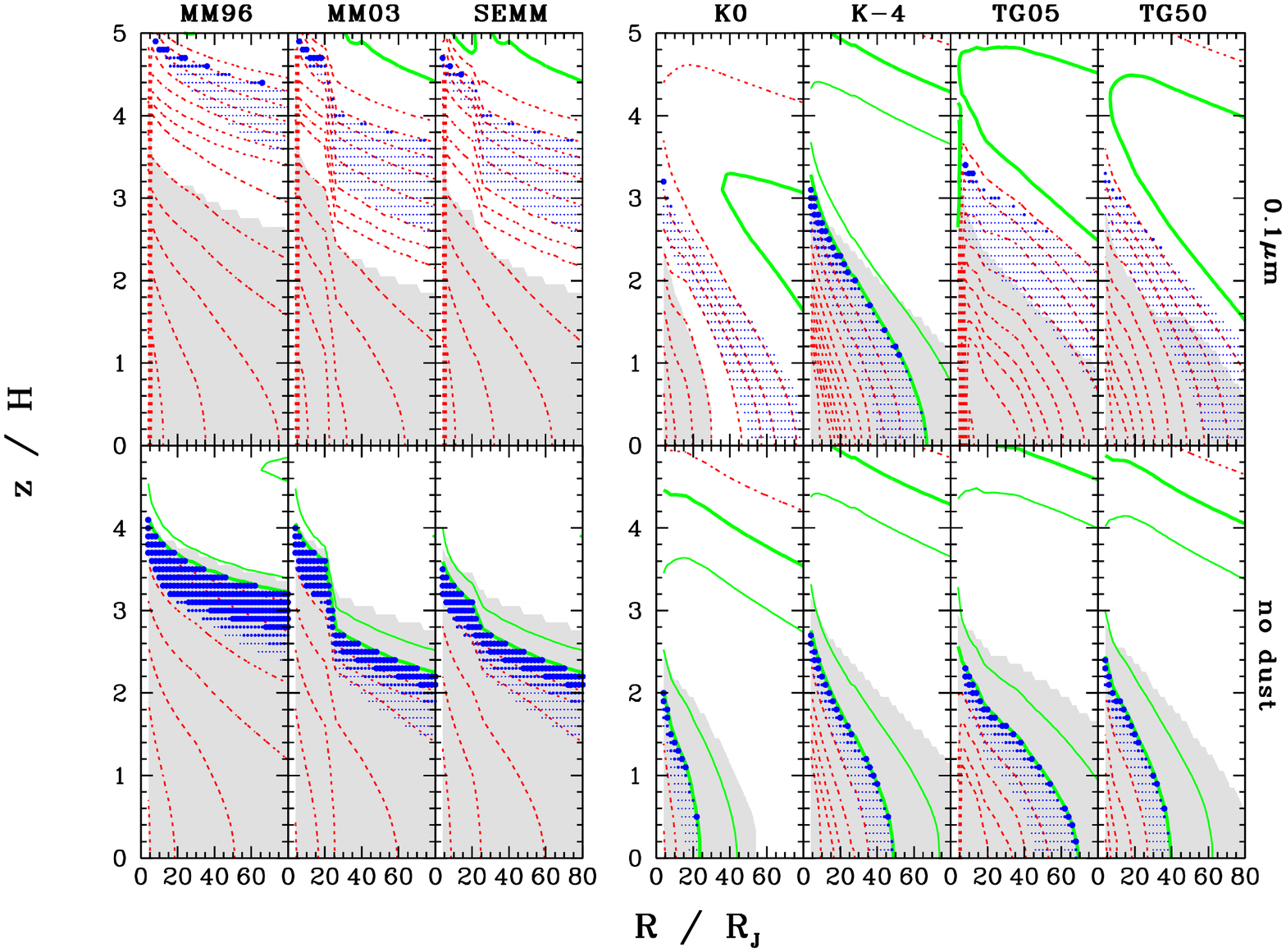}
\caption{\sf As figure~\ref{fig:deadxrcrsr} except that the cosmic ray
  ionization is omitted.  \label{fig:deadxrsr}}
\end{figure}

The seven circumjovian disk models from sections~\ref{sec:mmcjd}
and~\ref{sec:gssn} are shown together with their dust-free versions in
the next three figures.  Each figure corresponds to one ionization
scenario.  Figure~\ref{fig:deadxrcrsr} has X-rays, cosmic rays and
short-lived radionuclides.  Figure~\ref{fig:deadcrsr} has the X-rays
switched off.  In figure~\ref{fig:deadxrsr} the X-rays are restored,
while the cosmic rays are switched off.  The first of these three
scenarios is most favorable for magnetic activity, the second is
relevant if no protosolar X-rays reach the Solar nebula gap where
Jupiter and its disk reside \citep[as assumed
  by][]{2011ApJ...743...53F}, and the third covers the possibility
that the wind from the young Sun screens out the 0.1-1~GeV cosmic rays
contributing most to the ionization \citep{2011ApJ...735....8P}.  Each
figure has fourteen panels, corresponding to the seven model disks
with and without dust.  In figure~\ref{fig:deadxrcrsr} the left top
and bottom panels are identical to figure~\ref{fig:deadupclose2},
while the fifth panels in the top and bottom rows reiterate
figure~\ref{fig:deadupclose}.  Our calculations extend from 4$R_J$
inside Io's present orbit out to 80$R_J$ or a little more than 10\% of
the planet's Hill radius, and from the equatorial plane up to 5$H$.

In interpreting these figures, recall that X-rays ionize the uppermost
few tens of grams per square centimeter, the cosmic rays the next few
hundred and the radionuclides dominate in the deeper interior
(figures~\ref{fig:sd} and~\ref{fig:ioniz}).

Comparing figures~\ref{fig:deadxrcrsr} and~\ref{fig:deadcrsr} we see
that the X-ray ionization is crucial for magnetic activity.  With
X-rays, all gas-starved models show activity at least in their outer
reaches.  Without X-rays, none of the seven dusty models has a green
contour with the sole exception of TG05, where temperatures near
$4R_J$ are high enough for thermal ionization.  Removing the X-rays
greatly reduces the ion density near the disk surface, pushing down
the active layer's ambipolar-diffusion-dominated top boundary in many
cases so far that it meets the bottom boundary and the activity is cut
off completely.  Note that since the blue dots show turbulent mixing's
effects on the Ohmic resistivity, the dots are offset from the dead
zone edges in locations where those edges are determined by ambipolar
diffusion.  Mixing is effective only in locations where the blue dots
are adjacent to layers subject to MRI turbulence.

In contrast, the cosmic rays have little effect on the dead zone's
size.  Figure~\ref{fig:deadxrsr} differs from
figure~\ref{fig:deadxrcrsr} mostly in having more contours in the
disks' interiors, indicating the dead zones are more thoroughly dead.
As \cite{2011ApJ...743...53F} found, cosmic rays by themselves
generally provide too little ionization to support the MRI in the
circumjovian disk.  An exception is the minimum-mass models with no
dust.  Here the dead zone boundary lies near the cosmic ray
penetration depth, and is shifted upward a few tenths of a scale
height with the cosmic rays excluded.

We also experimented with replacing the short-lived radionuclides by
the less-ionizing long-lived radionuclides (not shown).  The
minimum-mass models' deep interiors are even more diffusive, but the
active layer boundaries, with ionization controlled by X-rays or
cosmic rays, are unaffected.

Considering the three figures~\ref{fig:deadxrcrsr}-\ref{fig:deadxrsr}
together, we see that in all cases with substantial activity
(i.e.\ where the Elsasser number exceeds ten somewhere) the active
layer's bottom boundary lies in or near the Ohmic-dominated region.
On the other hand, in all cases where the active layer has an upper
boundary lying on our grid, that boundary is set by ambipolar
diffusion.  The upper boundary in many cases also experiences slow MRI
growth because it lies above the heights of 3 and 3.7$H$ where the
total and vertical plasma beta fall to unity.

Again considering the three figures together, we see that local
chemical equilibrium is mostly a good approximation.  Mixing is
capable of changing the dead zone boundary by only a few tenths of a
scale height.  The thickest mixing layer occurs in the MM96 model
without dust, where X-ray ionization yields enough free electrons to
activate the MRI a half scale-height into the equilibrium dead zone.

Looking separately at the minimum-mass models, we see that all with
dust are quite dead.  Magnetic activity is possible only in a
low-surface-density zone outside the orbit of Callisto like that
advocated by \cite{2003Icar..163..198M}.  However the dust-free
versions of the three minimum-mass models all have more substantial
MRI-unstable upper atmospheres.  Far from the planet these even extend
below $3H$, where the gas pressure exceeds the magnetic pressure.

The situation is quite different in the gas-starved models, which have
an active layer in every scenario with X-rays.  The dead zone's size
varies among the dusty gas-starved models owing to the differing dust
abundances and surface densities.  The dusty K0 model is MRI-stable
near the planet due to ambipolar diffusion associated with its low
densities, while the dusty K-4 model's active layers comfortably reach
the planet if X-rays are included.  The TG50 model has a similar gas
surface density to the K-4, but a hundred times greater dust abundance
and thus a smaller active region.  The TG05 model, with its stronger
accretion heating due to a higher mass flow rate, is hot enough inside
$6R_J$ for collisional ionization in a surface layer.  The results are
otherwise insensitive to the details of the temperature and density
structure: the dead zones are little-changed when we make the
gas-starved disks vertically isothermal at the accretion temperature.

\subsection{Magnetic Field Strength}

Other choices for the magnetic field strength will change the picture
as follows.  The Ohmic diffusivity is independent of the field
strength, while the ambipolar diffusivity is generally proportional to
the magnetic pressure \citep{2007Ap&SS.311...35W}.  The Elsasser
number, which by eq.~\ref{eq:elsasser} depends on the ratio of the
magnetic pressure to the diffusivity, thus varies in proportion to the
magnetic pressure if the Ohmic term dominates, and is
field-strength-independent in the ambipolar regime.

As an example, if the vertical magnetic field has a pressure ten times
greater than we assumed above, reaching 1\% of the midplane gas
pressure, then (1) where the Ohmic term dominates, the Elsasser number
is an order of magnitude larger.  The circumplanetary disk's interior
is better-coupled.  (2) Where the ambipolar term dominates, the
Elsasser numbers are unchanged.  The active layer's top edge typically
does not move.  (3) The Ohmic-to-ambipolar transition shifts deeper by
about one contour.  In some cases the dead zone's lower boundary
switches from the Ohmic to the ambipolar regime.

%%%%%%%%%%%%%%%%%%%%%%%%%%%%%%%%%%%%%%%%%%%%%%%%%%%%%%%%%%%%%%%%%%%%%%%%%%%%%%%
\section{SOLID MATERIAL\label{sec:solids}}

In this section we discuss the implications of the circumjovian disk's
magnetic activity for the evolution of the solid material inside.  The
abundance of solids in the material delivered to the disk could be
less than suggested by Jupiter's overall heavy-element abundance of
about three times Solar \citep{2003NewAR..47....1Y}.  The nearby Solar
nebula may have become depleted in solids during the assembly of the
planet's core \citep{2005Icar..179..415H}.  Also, once the planet
grows massive enough to open a gap in the Solar nebula, only the
fraction of the solid mass contained in small grains is readily
carried to the vicinity of the planet according to hydrodynamical
results from \cite{2007A&A...462..355P}.  Grains smaller than about
10~$\mu$m are accreted on the planet and its circumplanetary disk, and
particles of intermediate size are captured by gas drag in the
pressure maxima immediately inside and outside the planet's orbit.
Large bodies become trapped in orbital resonances
\citep{1985Icar...62...16W}.

We saw above that solids in the form of fine dust particles can
prevent magneto-rotational turbulence through their large
recombination cross-section.  Removing the dust can restore turbulence
in the layers where the ionizing radiation is absorbed.  We would
therefore like to know, if turbulence is absent, how long before the
dust settles out?

\begin{figure}[tb!]
\epsscale{1.15}
\plottwo{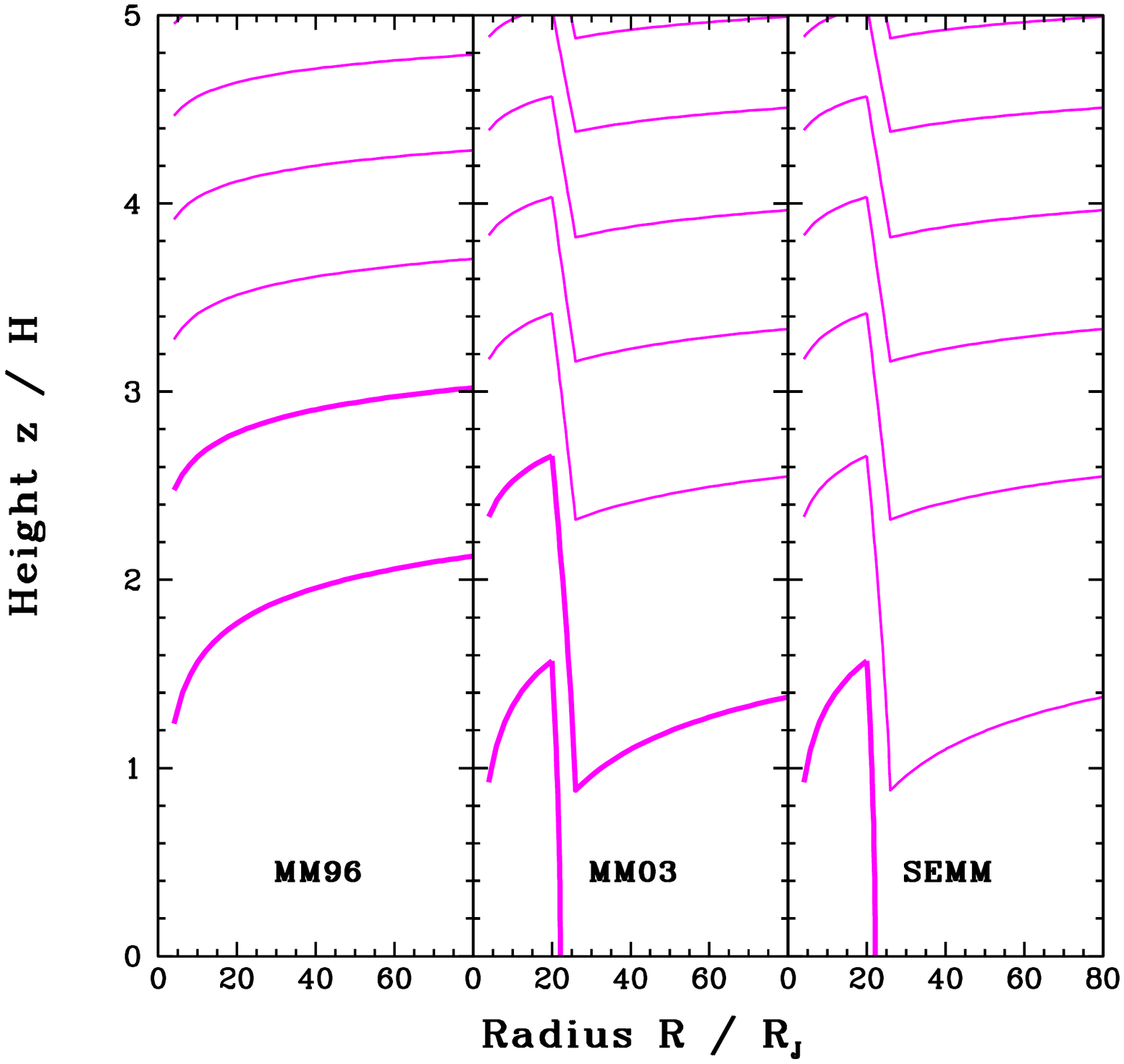}{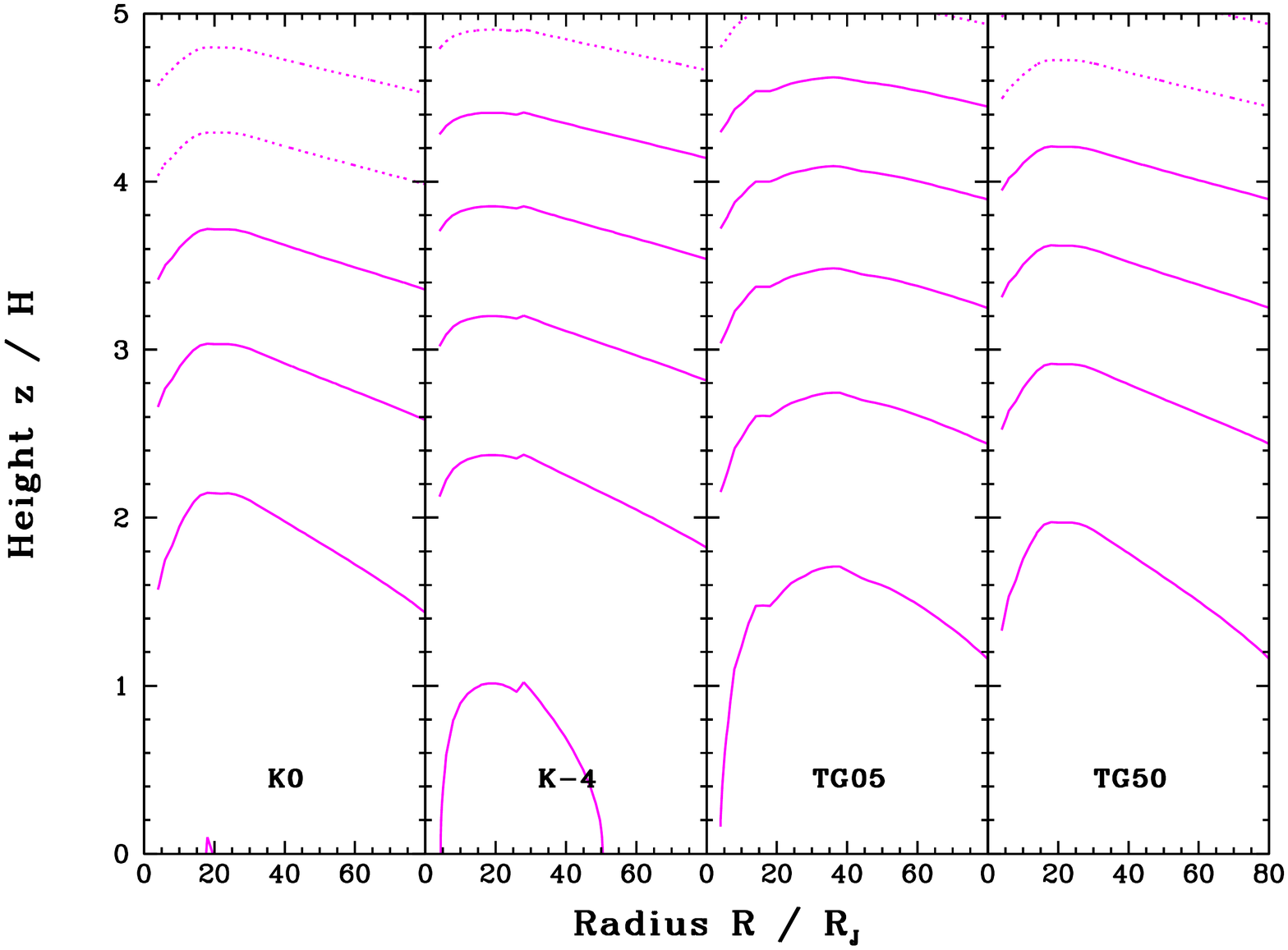}
\caption{\sf Settling time for 0.1-$\mu$m grains as functions of
  position in the three minimum-mass (left) and four gas-starved
  models (right).  The contours are logarithmic with spacing of one
  decade.  Settling times are 0.1~year and less on the dotted
  contours, 1 to $10^5$~years on the thin solid contours and a million
  years and longer on the heavy solid contours.  \label{fig:settle}}
\end{figure}

The settling time is the distance to the equatorial plane divided by
the grains' terminal speed.  In the Epstein regime, where the
particles are smaller than the gas molecules' mean free path and drift
through the gas slower than the sound speed, the settling time $t_S =
(\Omega^2 t_D)^{-1}$ where the gas drag stopping time $t_D =
(\rho_d/\rho)(a/c_s)$ \citep[e.g.][]{2010ApJ...708..188T}.  Recalling
that our grains are $a=0.1$~$\mu$m in radius with internal density
$\rho_d=2$~g~cm$^{-3}$ we obtain the timescales shown in
figure~\ref{fig:settle}.

In the minimum-mass models (left panels), which are stable against MRI
turbulence if dusty, the grains settle below 3$H$ in under a million
years.  Note that we have neglected coagulation.  Grains that stick
together to form compact aggregates on colliding settle still faster
\citep{2005A&A...434..971D}.  Since the minimum-mass models by
construction receive no resupply from outside, it seems likely the
atmosphere will become dust-depleted.  Good coupling to the fields
(figures~\ref{fig:deadxrcrsr}-\ref{fig:deadxrsr}) will then let some
of the atmosphere accrete on the planet, leaving the disk overall
gas-depleted.

In the gas-starved models (right panels), the settling is so fast that
the atmosphere would become dust-free down to $3H$ in less than a
thousand years --- except that in these models, dust and gas are
resupplied across much of the disk surface.  The atmosphere's dust
abundance will therefore be determined by external factors.  Near the
disk midplane, the dust is resupplied faster than it settles.  The
resupply time is the disk surface density divided by the incoming mass
flux.  Resupply is slowest at the inner boundary where its timescale
is under 2000~years in all four gas-starved models -- lower than the
midplane settling time in all cases.  Consider also that good magnetic
coupling extends to the midplane in the models' outermost annuli at
low dust abundances (figures~\ref{fig:deadxrcrsr}-\ref{fig:deadxrsr}).
MRI turbulence can then loft grains from the interior into the
atmosphere.  The most likely outcome for the gas-starved disks is thus
an atmosphere containing some dust overlying an interior with a higher
dust-to-gas ratio.

Both minimum-mass and gas-starved models contain layered dead zones
where the local MRI cannot drive turbulence.  The dead zones provide
favorable quiescent environments for growing larger bodies
\citep{2012MNRAS.422.1140G} which could eventually be assembled into
the regular satellites \citep{2013MNRAS.428.2668L}.

The Epstein regime applies throughout figure~\ref{fig:settle}, with
two minor exceptions.  First, the grains settle at a terminal speed
that exceeds the sound speed in a fraction of the topmost scale height
in the gas-starved models.  This simply means these low-density
regions quickly lose their dust.  Second, the gas mean free path is
less than the grain size at the midplane in the innermost annulus of
the MM96 model.  This densest point of all in the seven models is dead
with or without dust, so the settling time there is irrelevant for our
purposes.

%%%%%%%%%%%%%%%%%%%%%%%%%%%%%%%%%%%%%%%%%%%%%%%%%%%%%%%%%%%%%%%%%%%%%%%%%%%%%%%
\section{SUMMARY AND CONCLUSIONS\label{sec:conclusions}}

We examined the prospects for magnetic activity in seven different
models of the circumjovian disk, spanning a range of surface densities
and including both minimum-mass and gas-starved models.  The
gas-starved models were refined to include temperature-dependent
opacities and properly treat annuli of low optical depth.  For each
model we computed the ionization state, treating gas-phase
recombination, charge transfer to long-lived metal atoms, and
adsorption of the free charges on grains.  Where grains are present,
the last of these is the main recombination channel everywhere that
the ionization fraction is low enough for the grains to remain within
one or two electrons of neutral.

From the abundances of all the charged species we computed the
magnetic diffusivities, including the contributions from both Ohmic
diffusion and ambipolar drift.  The magnetic forces can drive
turbulence if the distance the field diffuses per orbit is less than
the fastest-growing wavelength of the MRI -- that is, if the
dimensionless Elsasser number $\Lambda = v_{Az}^2/(\eta\Omega)$ is
bigger than unity.  To see where magneto-rotational turbulence is
possible, we plotted contours of the Elsasser number vs.\ distance
from the planet and height above the equatorial plane, including both
Ohmic and ambipolar terms in the diffusivity $\eta$ and choosing a
magnetic field whose vertical component has a pressure 0.1\% of the
midplane gas pressure.  We investigated two limiting cases: the
stellar X-rays either (a) reach the circumplanetary disk with the same
flux as at the corresponding column in the nearby Solar nebula, or (b)
are blocked completely by the nebula.

All the dusty minimum-mass models we considered are thoroughly
magnetically dead, with or without X-ray ionization.  MRI turbulence
is unlikely to occur near the midplane in the dense, cold parts of a
minimum-mass circumjovian disk.  However if turbulence is absent, the
dust will settle out of the upper layers in a relatively short time.
Removing the dust greatly reduces the recombination rate, allowing MRI
turbulence in a surface layer reaching down to near the cosmic ray
penetration depth, or if cosmic rays are excluded, the X-ray
penetration depth.  In particular, surface layer angular momentum
transport by magnetic forces is possible in models where the surface
density falls off steeply beyond Callisto.  These models might need to
be modified to include stronger accretion stresses in the outer part.
More generally, if the dust-depleted surface layers accrete on the
planet or are removed through photoevaporation, the disk will be left
gas-poor overall.  It seems possible that minimum-mass models turn
into something resembling the solids-enhanced minimum-mass models put
forward by \cite{2003Icar..163..198M, 2003Icar..163..232M} and
\cite{2009euro.book...27E}.

By contrast, all the gas-starved models whether dusty or dust-free
have an accreting surface layer, with one group of exceptions: in the
dusty cases without X-rays, the ion densities are so low that
ambipolar diffusion prevents magnetic fields from acting on the bulk
neutral gas.  Another consequence of the gas-starved models' low
densities is that grains rapidly settle out in the absence of
turbulence.  A supply of fresh dust and gas from the Solar nebula is
assumed in constructing these models.  The resupply is fast enough to
keep the interior dusty, but too slow to prevent settling from
partially depleting the atmosphere.  Furthermore, we saw that
turbulence is capable of reaching the midplane in outer annuli, so
small grains can be returned to the atmosphere through mixing.  The
most likely outcome is an atmosphere with a reduced dust content.

In summary, both minimum-mass and gas-starved models of the
circumjovian disk have conductivities generally sufficient for
magnetic forces to provide the assumed accretion stresses.  However, a
key quantity is the X-ray flux reaching the neighborhood of the
planet.  Without the X-rays, the dusty gas-starved models couple to
magnetic fields too poorly for magneto-rotational turbulence to
operate.  The minimum-mass models' low internal conductivities with or
without X-rays, on the other hand, are as required to keep the
material in place.  Both classes of models can develop layered dead
zones, which could provide a favorable quiescent environment for
assembling regular satellites \citep{2013MNRAS.428.2668L}.

The magnetic coupling maps point to several more areas where our
understanding is lacking.  Future minimum-mass modeling may need to
treat the loss of dust-depleted gas from the surface layers.  In the
gas-starved models, the stress-to-pressure ratio ought to increase
with radius.  Consequently the material will pile up at locations
where the inflow slows.  Episodic accretion outbursts will result if
some additional angular momentum transport process switches on when
the disk surface density grows large enough.  The trigger can be the
gravitational instability if the accretion bottleneck fills up so much
that a gas-starved model approaches the surface densities of a
minimum-mass model \citep{2012ApJ...749L..37L}.

For the future development of the gas-starved models it is important
to address the issue of weak magnetic coupling in the absence of
stellar X-rays.  This motivates more careful calculations of the
transfer of the X-rays into the gap opened in the Solar nebula by
Jupiter's tides.  The X-rays could have reached the circumplanetary
disk at full strength if the Solar nebula interior to the planet's
orbit was cleared away, as appears to have happened in the so-called
transitional disks observed around some young stars today
\citep{2005ApJ...630L.185C, 2010ApJ...708.1107M, 2011ApJ...732...42A}.
Also, planets lying nearer their stars can have better-ionized disks
owing to the greater X-ray intensities.

Further constraints on the conditions in the circumjovian disk can
potentially be derived from the Laplace resonance.  The three inner
Galilean moons, Io, Europa and Ganymede, have orbital periods nearly
in the ratio 1:2:4.  \cite{2002Sci...298..593P},
\cite{2010ApJ...714.1052S} and \cite{2012ApJ...753...60O} demonstrated
that resonances can be assembled outside-in during satellite formation
and migration in gas-starved subnebula models.  Whether this works in
circumjovian disk models with magnetic stresses remains to be seen.

We have focused on the cold parts of the disk.  The thermally-ionized
zone near the planet could be important for its role in regulating the
planet's spin \citep{2011AJ....141...51L}, launching bipolar jets
\citep{2003A&A...411..623F, 2006ApJ...649L.129M} and determining
whether the planet begins its life cold or warm
\citep{2007ApJ...655..541M}.  The strong temperature dependence of the
ionization in this regime suggests that better thermodynamical models
are needed.

%%%%%%%%%%%%%%%%%%%%%%%%%%%%%%%%%%%%%%%%%%%%%%%%%%%%%%%%%%%%%%%%%%%%%%%%%%%%%%%
\begin{acknowledgments}
  This work was supported by the NASA Outer Planets Research program
  through grant 07-OPR07-0065, by the Hong Kong Research Grants
  Council through grant HKU 7024/08P, and by the Center for Planetary
  Science at Kobe University under the auspices of the MEXT Global COE
  program titled ``Foundation of International Center for Planetary
  Science.''  The work was carried out in part at the Jet Propulsion
  Laboratory, California Institute of Technology.  Copyright 2013.
  All rights reserved.
\end{acknowledgments}

%%%%%%%%%%%%%%%%%%%%%%%%%%%%%%%%%%%%%%%%%%%%%%%%%%%%%%%%%%%%%%%%%%%%%%%%%%%%%%%
\bibliographystyle{apj}
\bibliography{jmf}

\end{document}